\documentclass[aps,pra,twocolumn,groupedaddress,floatfix,showpacs]{revtex4-2}
\usepackage{amssymb,amsmath,amsthm}
\usepackage{mathtools}
\usepackage{graphicx}
\usepackage[english]{babel}
\usepackage{float}
\usepackage[svgnames]{xcolor}
\usepackage{blindtext}
\usepackage{bm}
\usepackage{comment}
\usepackage{etoolbox}
\usepackage{tikz,pgf}

\usepackage[symbol]{footmisc}

\usetikzlibrary{backgrounds,decorations.pathreplacing,arrows,decorations.pathmorphing,decorations.shapes,shapes.geometric,plotmarks}

 \newcommand{\ket}[1]{\left|#1\right\rangle}

\begin{document}

\title{Analytic Evolution for Complex Coupled Tight-Binding Models: \\ Applications to Quantum Light Manipulation}

 \author{Santiago Rojas-Rojas$^{1,2}$}
 \author{Camila Mu\~noz$^{1,3}$}
 \author{Edgar Barriga$^4$}
 \author{Pablo Solano$^{1,5}$}
 \author{Aldo Delgado$^{1,2}$}
  \author{Carla Hermann-Avigliano$^{2,3}$} 
\thanks{Corresponding author: santirojas@udec.cl}
\affiliation{$^1$ Departamento de F\'isica, Universidad de Concepci\'on, 160-C Concepci\'on, Chile}
\affiliation{$^2$Millennium Institute for Research in Optics (MIRO), Chile}
\affiliation{$^3$Departamento de F\'{\i}sica,  Facultad de Ciencias F\'isicas y Matem\'aticas, Universidad de Chile, Santiago, Chile}
\affiliation{$^4$Departamento de F\'{\i}sica, Facultad de Ciencias, Universidad de Chile, Santiago, Chile}
\affiliation{$^5$CIFAR Azrieli Global Scholars program, CIFAR, Toronto, Canada.}


\begin{abstract}
We present analytic solutions to the evolution in generalized tight-binding models, which consider complex first-neighbor couplings with equal amplitude and arbitrary phases. Our findings provide a powerful tool for efficiently calculating expectation values and correlations within the system, which are otherwise difficult to compute numerically. We apply our results to relevant examples in quantum light manipulation using N-port linear couplers, describing the evolution of single(multi)-mode squeezing, single photon added (subtracted) Gaussian states, and second-order site-to-site photon correlations. Significantly, our analytic results outperform standard numerical calculations. Our study paves the way for a comprehensive mathematical framework describing the spatial evolution of quantum states across a wide range of physical systems governed by the tight-binding model. 
\end{abstract}

\maketitle

\section{Introduction}



A good physical model captures the nature of a phenomenon through a simple mathematical description; however, simple descriptions do not always lead to simple solutions. The tight-binding (TB) model is an iconic example of this. Non-interacting particles hopping between adjacent sites describe the essence of the quantum behavior of electrons in solids, their transport properties, and band structure. Although initially intended to describe electrons in solids~\cite{AshcroftMermin}, the TB model characterizes analogous systems such as optical lattices with cold atoms \cite{Bloch,Lewenstein,andersonBECs}, light propagating in lattice waveguides \cite{fsSzameit,Longhi2009,Chen,Lederer}, phononic crystals \cite{Chen}, surface waves and topological insulators \cite{imura}, and quantum random walks \cite{Bromberg,Peruzzo}. Regardless of the simplicity of the TB model, many configurations lack analytic solutions in real space, and standard approaches resort to solving the problem in momentum space or through numerical methods. These approaches, however, pose a challenge for modeling the dynamics of spatial distributions and correlations of particles - critical aspects of quantum systems.

Although the TB model describes linear evolutions, its Hilbert space grows exponentially with the number of particles, presenting a significant computational problem \cite{Peruzzo}. For example, the seemingly simple scenario of a TB model with linear coupling between arbitrary sites 
lacks a real-space analytic solution to compute transport and correlations dynamics efficiently. Even approximated numerical methods with a truncated Hilbert space, mean-field descriptions \cite{Bloch}, or solving for the ground state \cite{exdiag} are resource-intensive tasks for classical computers \cite{qlineales}. More importantly, quantum phenomena generally arise from correlations beyond mean-field, while correlations between particles across lattice sites are not captured in the momentum-space representation that diagonalizes the TB Hamiltonian. Thus, we need techniques beyond the standard ones to calculate the real-space evolution of many-body quantum systems.

The previous limitations are especially relevant for multimode quantum light engineering, where correlations are crucial. For example, an N-port linear coupler described by a TB model can be used to produce multimode squeezed states \cite{nosotros}, besides the multiple applications it has for technologies based on integrated photonics \cite{intphotonics}. Furthermore, states generated by the addition or subtraction of photons on Gaussian states find an extension to multimode systems producing a plethora of non-classical states such as two-mode photon-added entangled coherent squeezed states \cite{Karimi}, (superpositions of) photon-added-trio-coherent-states \cite{Dat3modtrio}, and photon-added and photon-substracted four-mode squeezed vacuum states \cite{Karimi,YangYang}. Multimode quantum optics find applications in quantum technologies such as information processing \cite{WalschaersPRA}, metrology \cite{zhang2023,jiandong2018}, and cryptography \cite{expmultieng,Braun2014}, evidencing the relevance of calculating the evolution and the correlations of such states. Although real coupling coefficients describe the interactions in the above examples, many other models rely on complex coefficients \cite{Bender,mu2009,aharonov,abohmcaging}. For instance, complex phases of the coupling constants simulate a fictitious vector potential in honeycomb photonic lattices demonstrating photonic topological insulation \cite{szameit}, display an effective gauge potential for photons with synthetic dimensions in photonic lattices \cite{bell2017}, and lead to non-Hermitian Hamiltonians describing phenomena such as Bloch oscillations, the invisibility of defects, and $\mathcal{PT}$-symmetric quantum fields \cite{Longhi2016}.

This work presents an analytic real-space solution to the one-dimensional TB model for complex coupling constants with equal amplitudes and arbitrary phases. Our method draws upon tools from graph theory to analytically calculate all the elements of the transformation matrix for closed and open arrays, deriving expressions for the evolution of functions that depend on the annihilation and creation operators. The solution allows us to efficiently analyze transport dynamics and the propagation of spatial quantum correlations, outperforming numerical computations.  

We demonstrate the advantages of our findings by addressing problems related to quantum light propagation in an arbitrary N-port array of linear couplers. First, we elucidate the generation of multimode squeezing from single-mode squeezing. Second, we explore the dynamics of single-photon added or subtracted Gaussian states, providing a precise analytical depiction of their N-mode Wigner function, extending recent results \cite{WalschaersPRL} to the multimode regime. Third, we provide an analytic solution for propagating second-order photon correlations through coupling phase disorder. In particular, we show that 
two-photon entangled states exhibit an interference term in their second-order correlation that decreases exponentially with the degree of phase disorder, transitioning from a quantum to classical behavior. Finally, we benchmark our analytic solutions against standard numerical solvers proving their computational advantage. Our results enable the analysis of spatial evolution and correlation dynamics in an extended one-dimensional TB model with numerous particles. This approach broadens the scope of linearly coupled systems beyond traditional mean-field approximations and paves the way for exploring analytical solutions in varied spatial configurations.

\begin{figure*}[!t]
 \centering
\includegraphics[width=0.7\textwidth]{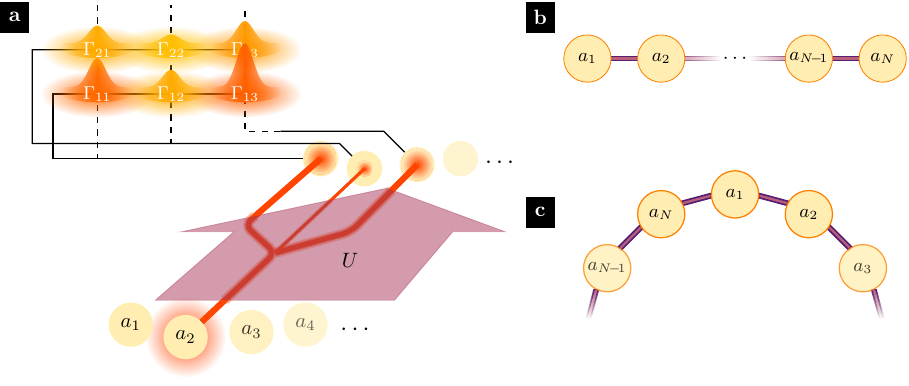}
 \caption{$N$-mode linear coupler. We consider a one-dimensional system sustaining $N$ modes of a bosonic field. The system evolves according to the transformation $U=\exp(-iHt)$ associated to the Hamiltonian in Eq. \eqref{eq:H}, allowing adjacent modes to couple according to a TB model with complex coefficients $C_j$. After evolution, we can analytically compute different observables or correlations $\Gamma_{ij}$.
 (c) and (b) show an schematic of an open and closed array respectively.}
\label{fg:modelo}
\end{figure*}

\section{Field modes transformations in a $N$-dimensional coupler}\label{sc:model}

We consider a system of $N$ coupled bosonic modes described by the Hamiltonian
\begin{equation}\label{eq:H}
 H=\hbar \sum\limits_{\langle j,k\rangle}\left(C_{j k}a_j^\dagger a_k+C_{jk}^\ast a_k^\dagger a_j\right)\,,
\end{equation}
where $a_j$ ($a_j^\dagger$) is the annihilation (creation) operator of an excitation in the $j$-th mode. In the most general case, the sum runs over all the ordered pairs of modes, whose coupling is determined by the respective complex coefficient $C_{jk}$. In the context of the examples given in this work, this Hamiltonian describes a linear device with $N\times N$ input-output ports. We refer to the system as coupler, array, or $N$-mer interchangeably.



We use a Heisenberg picture approach for $N$-bosonic modes and apply it to the open and closed TB model with complex coupling coefficients (see Fig. \ref{fg:modelo}). The evolution of each mode is computed as a transformation of the bosonic operators $a_m$ into new bosonic operators $a_m'$ given by
\begin{equation}
a_m'=U^{\dagger} a_m U,
\label{eq:UaU}
\end{equation}
where $U=\exp(-iHt/\hbar)$ is the evolution operator. This input-output approach is suitable for studying the spatial evolution and correlations of arbitrary states in a $N$-mode system, as it is state-independent. 

The Baker-Campbell-Hausdorff (BCH) formula allows us to express the transformed operators $a_m'$ as
\begin{equation}\label{eq:expBCH}
 a_m'=a_m+\left[\frac{i}{\hbar} t H, a_m\right]
 +\frac{1}{2!}\left[\frac{i}{\hbar} t H,\, \left[\frac{i}{\hbar} t H, a_m\right]\right]+\cdots\,,
\end{equation}
or equivalently
\begin{equation}\label{eq:expBCHb}
a_m'=\sum_{n=0}^\infty\frac{1}{n!}\left(\frac{it}{\hbar }\right)^n\left[H,a_m\right]_n\,,
\end{equation}
with
\begin{equation}\label{eq:conmHam}
 \left[H,a_m\right]_n=\left[H,\left[H,a_m\right]_{n-1}\right]=\sum_{j=1}^N\Lambda^{(n)}_ja_j\,,
\end{equation}
and $\left[H,a_m\right]_0=a_m$. The right-hand side arises from the canonical commutation relations and implies that each term in the iterative sum in Eq.~\eqref{eq:expBCHb} is a linear combination of the $N$ modes of the field. Substituting with the definition of $H$ in Eq.~\eqref{eq:H} we obtain
\begin{equation}\label{eq:alfan}
    \begin{split}
        \Lambda^{(n+1)}_j &= -\hbar\sum_{\ell(j)} C_{\ell(j),j}\Lambda^{(n)}_{\ell(j)},
    \end{split}
\end{equation}
where the sum over $\ell$ is restricted to the modes connected to the $j$-th mode through the coupling constants. Thus, the set of coefficients $\Lambda_n^{(n)}$ and $\Lambda_n^{(n+1)}$ are related through the hermitian \emph{coupling matrix} $\mathcal C$, 
which corresponds to the single-particle representation of the Hamiltonian in Eq.~\eqref{eq:H}. Using Eq.~\eqref{eq:alfan} we can write Eq.~\eqref{eq:expBCHb} as
\begin{equation}\label{eq:expBCHc}
    a^{\prime}_m = \sum_{n=0}^{\infty}\frac{1}{n!}\left(\frac{it}{\hbar}\right)^n\left[\sum_{j=1}^N \hat e_j^\intercal \left(-\hbar\mathcal{C}^\intercal\right)^n\hat e_m a_j\right]\,,
\end{equation}
where $\hat e_j$ denotes the $j$-th vector in the canonical basis. The information specific to the array is present in each term through the coupling matrix $\mathcal{C}$. Writing the modes operators as the components of vectors $\vec{a}'$ and $\vec{a}$, the above expression can be written as
\begin{equation}
  \vec{a}'=A^{(N)}\vec{a},
    \label{MatrixEqA}
\end{equation}
where the $N\times N$ matrix $A^{(N)}$ is the exponential of the coupling matrix $\mathcal{C}$ (see Appendix \ref{Appendix:coupling_matrix} for details), that is, 
\begin{equation}\label{eq:expCoupling}
    A^{(N)}=\exp(-it\mathcal{C}).
\end{equation}
Thereby, determining the evolution of functions of bosonic operators depends on evaluating the transformation matrix $A^{(N)}$.

The dependence of operators $a_i$ on the dynamical variable $t$ provides a convenient framework for describing the evolution of specific observables, such as the mean photon number or the field quadratures \cite{Rai,Rostami} as well as classical and quantum correlations on few-particle systems \cite{Peruzzo,Bromberg}. However, increasing the number of particles presents a significant difficulty since the state of the system is nonseparable upon evolution, leading to an exponential increase in the dimensions required to describe the dynamics (see the graphical representation in Ref. \cite{Peruzzo}). The exponential growth of the dimensionality with $N$ also arises in approaches based on the Schr\"odinger picture \cite{RodriguezLara}. This challenge presents the need for alternative methods whose convenience depends on the particular system.

\subsection{Approaches to determine $A^{(N)}$}\label{sc:A}

\begin{figure}[!tb]
 \includegraphics[]{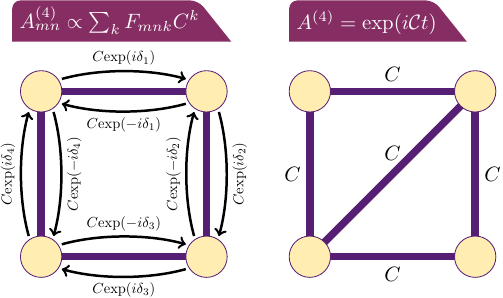}
 \caption{
Example of two similar linear couplers whose evolutions are easier to solve with different methods. For the closed tetramer with complex coupling (left), it is challenging to derive analytical expressions from $\exp(i{\mathcal C}t)$, but the evolution of the bosonic operators can be readily represented as a series expansion in terms of the amplitude $C$ (see Sec. \ref{sc:tb}). For the rhomboidal array (right) 
the exponential $\exp(i{\mathcal C}t)$ can be calculated analytically for real coupling coefficients. 
 }\label{fg:ejemplos}
\end{figure}

The problem reduces to finding the entries $A_{m,n}^{(N)}$ of the matrix $A^{(N)}$, which are particular to the geometry defining the interacting modes, i.e., the set of pairs in the sum in Eq.~\eqref{eq:H}. The first approach would be to diagonalize the Hamiltonian. However, one can only do so analytically in particular cases \cite{Rai}. Other methods allow obtaining an explicit expression for the $A_{m,n}^{(N)}$, whose suitability differs for each specific coupling matrix $\mathcal{C}$. For example, as discussed in Appendix  \ref{Appendix:Derivation}, by grouping the terms as factors of each mode $a_m$, the expansion in Eq.~\eqref{eq:expBCH} indicates that each entry $A_{m,n}^{(N)}$ is, in fact, expressible as a series expansion on the amplitude of the coupling coefficients $C_{jk}$, whose terms follow an integer sequence. The iterative relation between successive terms can be derived from known integer sequences or a generating function \cite{generatriz} specific for each mode coupling configuration. Indeed, this approach provides an expression for the entries $A_{m,n}^{(N)}$ but translates the problem into finding the initial condition for the respective sequence in their expansion. We find the previous approach based on integer sequences more convenient for studying the closed TB model with complex coupling coefficients. At the same time, direct evaluation of the exponential in Eq.~\eqref{eq:expCoupling} is advantageous for specific arrays. We present two illustrative examples in Fig. \ref{fg:ejemplos}.

On the other hand, the coupling matrix $\mathcal C$ links to graph theory since it is the adjacency matrix of the connected graph where the propagating particles describe a quantum random walk (QRW) \cite{Bromberg,Peruzzo}. This framework relates the system to an underlying support graph \cite{lu2016}, where each mode corresponds to a vertex while their couplings are the weighted edges. This correspondence includes the case of complex coupling coefficients, meaning complex entries of the adjacency matrix, which leads to asymmetric transport across the edges of the graph \cite{complex2013,complex2019} in the context of quantum random walks. Specifically, the dynamics in the $N$-mer is related to the problem of counting paths in a connected graph \cite{Znumbers}. From Eq.~\eqref{eq:expBCHc}, one sees that the expression enclosed in square brackets is formally equivalent to a sum over all the walks of length $n$ leading to vertex $m$. Thus, Eq.~\eqref{eq:expBCHc} implies that the transformed mode $a'_m$ results from adding the weights of all the paths leading to the $m$-th vertex from the vertices associated with the modes $a_j$. This fact will be helpful to derive the explicit form of the transformations in Eq.~\eqref{eq:UaU} for the open and closed TB couplers in the next section.

\subsection{Complex-coupling tight-binding model}\label{sc:tb}

We now focus on the $N$-mode TB model with first-neighbor coupling. For the open case in Fig. \ref{fg:modelo}(b), the Hamiltonian in Eq.~\eqref{eq:H} becomes
\begin{equation}\label{eq:Htb}
 H=\hbar \sum\limits_{j=1}^{N-1}\left(C_j a_j^\dagger a_{j+1}+C_j^\ast a_{j+1}^\dagger a_j\right)\,,
\end{equation}
where $C_j=C \exp(i\delta_j)$ is the complex coupling coefficient between the $j$-th and the $(j+1)$-th modes, being the amplitude $C$ and the phases $\delta_j$ real quantities. Eq.~\eqref{eq:Htb} also describes the closed array shown in Fig.~\ref{fg:modelo} (c) by extending the sum to $j=N$ and taking $a_{N+1}=a_1$. Most physical descriptions using Eq.~\eqref{eq:Htb} consider real and identical coupling constants, i.e. $\delta_j=0$, which implicitly assume time-reversal symmetry of the dynamics \cite{complex2013}. Here, we study a more general case exhibiting complex coupling coefficients with identical coupling amplitude between modes but different arbitrary phases $\delta_j$. 
Note that the complex coupling terms in Eq.~\eqref{eq:Htb} are still Hermitian, so its dynamics differ from the non-Hermitian TB networks \cite{Longhi2016}. Instead, it exhibits asymmetrical coupling between connected nodes of the support graph, which gives rise to time-reversal symmetry breaking \cite{Liu} and chiral quantum walks \cite{complex2013,complex2019}, whose actual effects strongly depend on the geometry and the parity of $N$ \cite{lu2016}.


The complex entries of the transformation matrix can be generally expressed as
\begin{equation}\label{eq:A_mn}
 A^{(N)}_{m,n}=\exp(-i\phi_{m,n})\beta^{(N)}_{m,n}.
\end{equation}
While this representation is valid for any complex value of the coefficients, it is not always possible to express it in an explicit form suitable for actual computations. 

For the open array (Fig. \ref{fg:modelo}(b)), an evaluation of the recursive terms in Eq.~(\ref{eq:conmHam}) leads to
\begin{equation}\label{eq:definicionPhi}
 \phi_{m,n}=\Delta_{m-1}-\Delta_{n-1}\,,
\end{equation}
with $\Delta_k=\delta_1+\delta_2+\cdots+\delta_k$ and $\Delta_0=0$. The $m,n$-th entry of the transformation matrix is found to have the general form
\begin{equation}\label{eq:Amn}
 \begin{split}
 A_{m,n}^{(N)}=& \frac{2}{N+1}\exp(-i[\Delta_{m-1}-\Delta_{n-1}])\\
 &\times\sum_{k=1}^N \exp\left(-2i\cos\left[\frac{k\pi}{N+1}\right]C t\right)\\
 &\quad \times\sin\left(\frac{mk\pi}{N+1}\right)\sin\left(\frac{nk\pi}{N+1}\right)\,.
 \end{split}
\end{equation} 


This expression can be derived from different approaches, discussed in Appendix \ref{Appendix:Derivation}.

For the closed array (Fig. \ref{fg:modelo}(c)), the translational symmetry of this case allows us to consider any mode as the first one. Then, we can deduce the structure of the transformation matrix $\overline{A}^{(N)}$, whose $n$-th row has the same entries as the first but \emph{cycled} by $n$ positions:
\begin{equation}
 \overline{A}^{(N)}=
 \begin{pmatrix}
   \overline{A}^{(N)}_{11} &  \overline{A}^{(N)}_{12} & \cdots & \overline{A}^{(N)}_{1,N-1}  & \overline{A}^{(N)}_{1N} \\
   \overline{A}^{(N)}_{1N} &  \overline{A}^{(N)}_{11} & \cdots &  \overline{A}^{(N)}_{1,N-2} &  \overline{A}^{(N)}_{1,N-1} \\
   \vdots &  \vdots & \ddots & \vdots &\vdots \\
   \overline{A}^{(N)}_{1,2} &  \overline{A}^{(N)}_{13} & \cdots & \overline{A}^{(N)}_{1N} & \overline{A}^{(N)}_{1,1} 
 \end{pmatrix}\,.
 \end{equation}
Therefore, only the entries $\overline{A}^{(N)}_{1,n}$ are required to obtain the full matrix. Unlike the open $N$-mer, the matrix elements for the closed array are divided into several cases. For $N$ even and $n$ odd, with the condition $n=N/2+1$, we found
\begin{equation}\label{eq:A_1n}
     \begin{split}
         \overline{A}_{1,n}^{(N)} &= 2\sum_{l=0}^{\infty} \cos{\left(\frac{N}{2}(2l+1)(\delta-\pi/2)\right)}J_{N(2l+1)/2}(2Ct)\,, 
     \end{split}
 \end{equation}
where the $J_{N(2l+1)/2}$ are the Bessel function of first kind and $\delta=\delta_j$ (same phase for every coupling). The remaining cases are listed in Eq.~\eqref{eq_appendix:A_1n} of the appendix \ref{Appendix:Successive_method}. Circulant matrices, such as $\overline{A}^{(N)}$, are diagonalized by means of the discrete Fourier transform and, consequently, linear functions of them can be efficiently evaluated by a fast Fourier transform.

It is important to highlight the versatility of the previous formalism, since Eq.~\eqref{eq:A_mn} 
holds for any set of coupling coefficients, beyond the two one-dimensional cases studied here. This allows us to extend our study to two-dimensional systems that can be mapped to a one-dimensional matrix, provided that the appropriate integer sequence (or its recurrence matrix) is known. Remarkable examples are \emph{flat-band lattices} whose dynamics can be addressed using a one-dimensional linear coupler \cite{moralesvicencio,santiagorojas}.

\begin{figure}[ht!]
 \centering
 \includegraphics[width=0.48\textwidth]{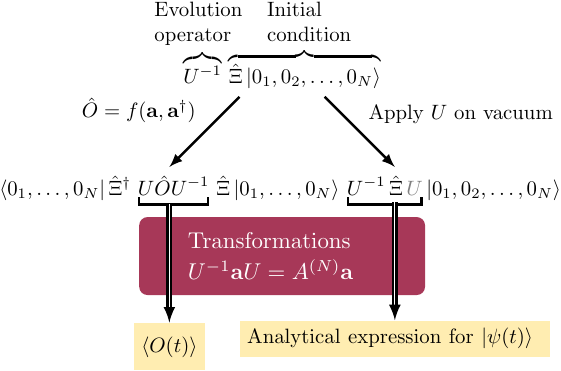}
 \caption{Schematic of the method to compute the evolution of the expectation value of any observable expressible as an analytic function of $\bf a$ and ${\bf a}^\dagger$ (left) or the state of the system (right). Knowing the analytical form of the transformation matrix $A^{(N)}$, we can consider any initial state generated by an operator $\hat{\Xi}$ composed of sums and products of the operators $a_j^\dagger$.}\label{fg:metodo}
\end{figure}

\section{Applications}\label{sc:qtc}

We use our previous results to describe the quantum evolution of many particles or excitations in $N$ linearly coupled modes. We aim to quantify quantum correlations between particles through their evolution (both in the position and the quadrature space). While the time required to solve the single-particle problem scales linearly with the number of modes $N$, it grows exponentially with the maximum number of particles. Computing the evolution of such systems is generally hard. However, it can be significantly simplified using the expressions obtained in the previous section.

\subsection{Method summary}\label{sc:method}

Figure \ref{fg:metodo} summarizes the method we follow to compute efficiently the evolution of states or expectation values. We express the initial state as an operator $\hat{\Xi}$ acting over the vacuum state, where $\hat{\Xi}$ is written as a combination of terms containing powers of the creation and annihilation operators. For example, in Section \ref{sc:squeezing}, we study the evolution from a product of single-mode squeezed vacuum states $\ket{\xi_1}\ket{\xi_2}\cdots\ket{\xi_N}$, so the corresponding operator will be $\hat{\Xi}=\hat{S}_1(\xi_1)\hat{S}_2(\xi_2)\cdots \hat{S}_{N}(\xi_N)$, being $S_j(\xi_j)$ the squeezing operator on the mode $j$ with squeezing parameter $\xi_j$. On the other hand, as a part of our study of two-particle correlations in Section \ref{sc:corrs} we require an initial state given by the superposition of Fock states $1/\sqrt{2}(\ket{2_k}+\ket{2_\ell})$ so we use $\hat{\Xi}=1/2[(a_k^\dagger)^2)+(a_\ell^\dagger)^2)]$.

The evolution is calculated through $U=\rm{exp}(-i\hbar H t)$. Depending on our goal, we choose between two different procedures. To obtain the state evolution, we operate with $U$ on the vacuum state, using the fact that it remains unaffected by rotations, obtaining $U^{-1}\hat{\Xi}U$. Since the operator $\hat{\Xi}$ can be decomposed in sums and powers of the creation operators $a_i$, we can then use the identity
\begin{equation}
 U^{-1}a^{\dagger}_jU=\sum_i A^{(N)\ast}_{ji} a^\dagger_i\,.
\end{equation}
and then apply our main results, summarized in Appendix \ref{Appendix:Main results} and deduced in Appendix \ref{Appendix:Derivation}. Specifically, we use Eq. \eqref{eq:A_mn}
for open arrays and Eq. \eqref{eq:A_1n} for the closed arrays. An arbitrary case can always be addressed by the exponential expression in Eq. \eqref{eq:expCoupling}
which, as discussed in Section \ref{sc:A}, is not the optimal method in most cases. On the other hand, to compute the expectation value of an observable $\hat{O}$ expressible as an analytic function of $a$ and $a^\dagger$, the transformations $Ua_{j} U^{-1}$ and $Ua^\dagger_jU^{-1}$ allow to evaluate $U\hat{O}U^{-1}$. This method is versatile enough to address evolution from a wide variety of initial conditions, which could be a superposition of Fock states, squeezed-coherent states, and even photon-added (subtracted) squeezed-coherent states \cite{Zavatta,TMPADSS,subTMSVS,Karimi,EXPsubTMSVS,Dufour17,highaddtriocoh}.

\subsection{Single-mode squeezing cancellation}\label{sc:squeezing}

\begin{figure}[!tb]
\begin{center}
\includegraphics[width=.48\textwidth]{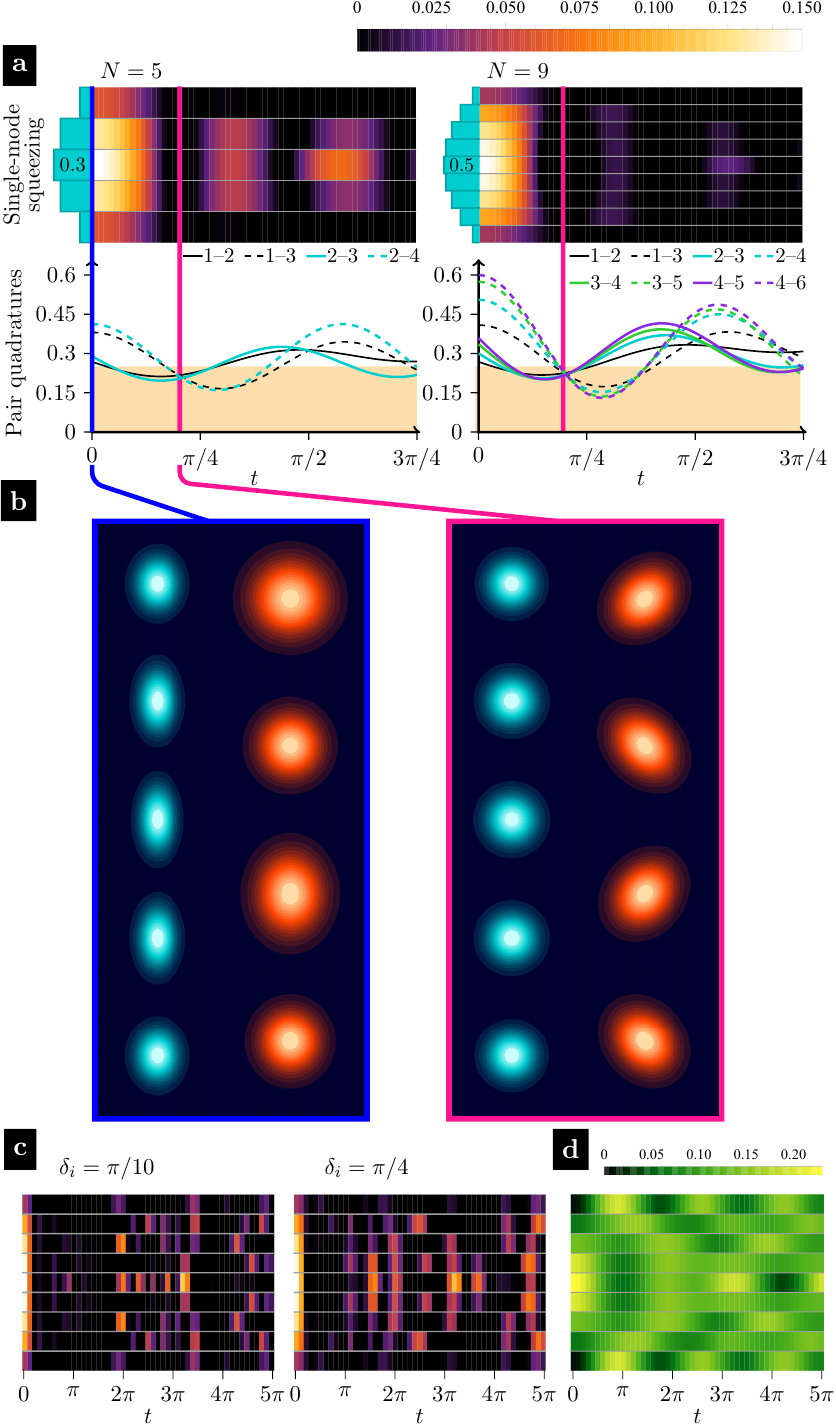}
\end{center}
 \caption{Transition from single-mode to multimode squeezing in linear couplers. (a) Evolution of the single-mode squeezing degree for the pentamer (top left) and the enneamer (top right). Black regions indicate the absence of single-mode squeezing. The relation between the respective squeezing parameters $\xi_i$ for each mode of the initial state is depicted by the horizontal blue bars, where we indicate the initial value at the center mode used in the simulations. The dynamics of the variance of two-mode quadratures for different pairs of modes is shown in the plots below ($N=5$ down left and $N=9$ down right). The orange region indicates squeezing. Vertical pink lines mark a distance where the state of the system exhibits multimode squeezing only. (b) Single-mode Wigner function (blue surfaces) and marginal distributions of the multimode Wigner function for the same pairs of quadratures (red surfaces) for $N=5$. (c) Examples of the evolution of the squeezing degree in the enneamer when complex coefficients with identical phases $\delta_i\neq 0$ are considered. (d) The particle number at each mode, here shown for the $N=9$ case, is the same for any value of the coupling phases. }\label{fg:squeezing}
\end{figure}

The propagation of squeezed states through an open $N$-mer has been previously studied \cite{Rai}. In particular, it has been shown that for $N=2$ and $N=3$, an initial product of single-mode squeezed states can evolve into a state exhibiting multimode squeezing with suppressed single-mode squeezing \cite{jmo,nosotros}. This procedure allows to induce multimode correlations among quantum fluctuations of different pairs of modes from a single-mode squeezed state. We use the mathematical tools developed here to generalize this phenomenon to an arbitrary $N>3$. Let us consider that the initial state of the system is given by a product of single-mode squeezed vacuum states:
\begin{equation}
 \ket{\psi_0}=\hat{S}_1(\xi_1) \hat{S}_2(\xi_2)\cdots \hat{S}_N(\xi_N)\ket{0}\,,
\end{equation}
being $\hat{S}_j(\xi_j)=\exp([\xi_j^\ast a_j^2 - \xi_j{a_j^\dagger}^2]/2)$ the squeezing operator on the $j$-th mode. As depicted in the diagram of Fig.~\ref{fg:metodo}, the state of the system after a time $t$ is obtained by applying the unitary operator $U^{-1}$ on the squeezing operators of the initial state and the inverse $U$ on the vacuum, leading to
\begin{equation}\label{eq:psit}
    \begin{split}
        \ket{\psi_t}=&\exp\Bigg\lbrace\frac{1}{2}\Bigg[\sum_{j=1}^{N}a_{j}^{2}\sum_{i=1}^{N}\xi_{i}^{*}\left(A_{ij}^{(N)}\right)^{2}\\
        &\qquad+2\sum_{k=1}^{N-1}\sum_{j=k+1}^{N}a_{k}a_{j}\sum_{i=1}^{N}\xi_{i}^{*}A_{ik}^{(N)}A_{ij}^{(N)}\\
        &\qquad\qquad-H.C.\Bigg]\Bigg\rbrace\ket{0_{1},\ldots,0_{N}}\,,
    \end{split}
\end{equation}
where the dependence on time $t$ is implicit in the coefficients $A_{ij}^{(N)}$ of the transformation matrix obtained in Section \ref{sc:tb}. From the first term in this expression, it follows that a single-mode squeezing suppression fulfill the condition:
\begin{equation}\label{eq:ccond}
   \sum_{i=1}^{N}\xi_{i}^{*}\left(A_{ij}^{(N)}\right)^{2}=0\,,\qquad \forall~j=1,2,\ldots,N\,.
\end{equation}

Let us consider real initial squeezing parameters. This assumption leads to a system of $N$ equations and $N+1$ unknowns: the $N$ initial squeezing parameters and the propagation distance $t$ at which the single-mode squeezing nullifies. However, since we are not interested in the exact amount of initial squeezing but in the relation between the squeezing parameters for all the modes, one squeezing parameter can be arbitrary, rendering the problem solvable. Furthermore, we can set the required initial distribution of squeezing to be symmetrical across the $N$-mer, i.e. squeezing in the $j$-th and $N+1-j$-th modes should be the same. This way, Eq.~\eqref{eq:ccond} leads to a system of $(N+1)/2$ equations ($N/2$ equations)  for odd $N$ (even $N$), with the same number of unknowns. Note that due to linear independence of Eq. \eqref{eq:ccond} and the columns of the transformation matrix, the sums $\sum_{i}\xi_iA_{ik}^{(N)}A_{ij}^{(N)}$ in the second term of Eq.~\eqref{eq:psit} will not be null in general for any pair $k,j$ when evaluated in the solutions $\xi_i^\ast$. Therefore, we can expect to find two-mode squeezing over different pairs of modes throughout the evolution where single-mode squeezing is suppressed. 

We exemplify this with a pentamer (N=5) with real first-neighbor coupling. By solving Eq.~\eqref{eq:ccond}, we find that the squeezing values $\xi_1=\xi_5=0.1$, $\xi_2=\xi_4=0.25$ and $\xi_3=0.3$ suppress single-mode squeezing at $t=0.640$ but not multimode-squeezing. Figure ~\ref{fg:squeezing} shows the squeezing dynamics for this case, along with analogous results for the enneamer ($N=9$). At certain stages of the evolution, single-mode squeezing vanishes while squeezing is observed in the variance of the two-mode quadratures between multiple pairs of modes. We confirm this by computing the single-mode and multimode Wigner functions at different values of $t$, shown for $N=5$ in Fig.~\ref{fg:squeezing} (b). The suppression of single-mode squeezing is clearly observed in the Wigner function of each mode (blue surfaces). Conversely, the marginal distributions of the multimode Wigner functions \cite{nosotros,Braunstein} become squeezed (red surfaces). 

Our results generalize previous ones, showing that $N$-mers can produce multimode squeezed states among arbitrary $N>3$ modes from initially single-mode squeezed states.  In addition, we computed the evolution of the squeezing degree for complex coupling coefficients, i.e. $\delta_i\neq 0$. As shown by the examples in Fig. \ref{fg:squeezing} (c), single-mode squeezing cancellation, with the corresponding transition to multimode squeezing (not shown in the examples), still occurs for the phases studied, with the particular evolution of the squeezing degree being different for each case. Interestingly, the effect of the complex coupling is not reflected in the evolution of the particle number [presented in Fig. \ref{fg:squeezing} (d) for the enneamer] which is the same for all the values of $\delta_i$ under consideration.

\subsection{Wigner function representation for propagating non-classical states}\label{sc:W}

Wigner functions are informationally complete graphic representations of the quantum states of single and multimode electromagnetic fields. Negative values in the Wigner function imply that field is purely quantum without a classical counterpart \cite{Kenfack2004,Siyouri2016}. Since its dynamics reflect the state evolution, the Wigner representation is useful for readily identifying emerging or vanishing quantum features \cite{Uria2020,Uria2023}, an otherwise challenging mathematical task. The results presented in Section \ref{sc:model} are particularly useful to track the variation of the multimode Wigner function describing non-classical states of the system. 

As an example, we study the propagation of initial states obtained by adding (subtracting) a single excitation to (from) a non-displaced Gaussian state and examine the Wigner function dynamics. In recent articles, a rather simple expression for the Wigner function $W_G(Y)$ of a single photon-added and substracted Gaussian state was presented \cite{WalschaersPRL,WalschaersPRA}, which uses the symplectic representation of the phase space. In this framework, we associate the in-phase and out-of-phase components of the $\ell$-th mode amplitude with two elements, $e^{(\ell)}$ and $Je^{(\ell)}$, in a \emph{symplectic basis}:
\begin{equation}
 {\mathcal E}^s=\left\lbrace e^{(1)},\ldots,e^{(N)},Je^{(1)},\ldots,Je^{(N)}\right\rbrace\,,
\end{equation}
where $J$ is a $2N\times 2N$ matrix fulfilling $J^2=-{\mathbb I}$ and $f_1 \cdot Jf_2=-f_2 \cdot Jf_1$ for any $f_1$ and $f_2$ linear combinations of the symplectic basis in ${\mathbb R}^{2N}$, sometimes called the \emph{symplectic form}. The operators that annihilate and create an excitation in the mode $f$ are denoted by $a(f)$ and $a^\dagger(f)$ and can be defined in terms of quadrature operators as
\begin{equation}
\begin{split}
a(f)&=\left(Q(f)-iQ(Jf)\right)/2\,,\\ a^\dagger(f)&=\left(Q(f)+iQ(Jf)\right)/2\,.
\end{split}
\end{equation}

In order to use our expressions for the transformation matrix $A^{(N)}$ with the symplectic formalism, we set the correspondence $a_j=a(e_j)$. Since the quadrature operators are linear, i.e., $Q(\alpha f_1+\beta f_2)=\alpha Q(f_1)+\beta Q(f_2)$, and $a(Jf)=-ia(f)$, we can describe the evolution $a_j'=Ua_jU^{-1}$ as a transformation of the corresponding mode in the symplectic space:
\begin{eqnarray}\label{eq:apsimp}
 a_j'&=&\sum_{\ell=1}^NA^{(N)}_{j,\ell} a_\ell =a\big(\sum_{\ell=1}^{N} ({\rm Re}\lbrace A^{(N)}_{j\ell}\rbrace e_{\ell}+{\rm Im}\lbrace A^{(N)}_{j\ell}\rbrace Je_{\ell})\big)\,,\nonumber\\
 a_j^{\dagger\prime}&=&a^\dagger\big(\sum_{\ell=1}^{N} ({\rm Re}\lbrace A^{(N)}_{j\ell}\rbrace e_{\ell}-{\rm Im}\lbrace A^{(N)}_{j\ell}\rbrace Je_{\ell})\big)\,.
\end{eqnarray}

\begin{figure}[ht!]
 \centering
\includegraphics[height=.67\textheight]{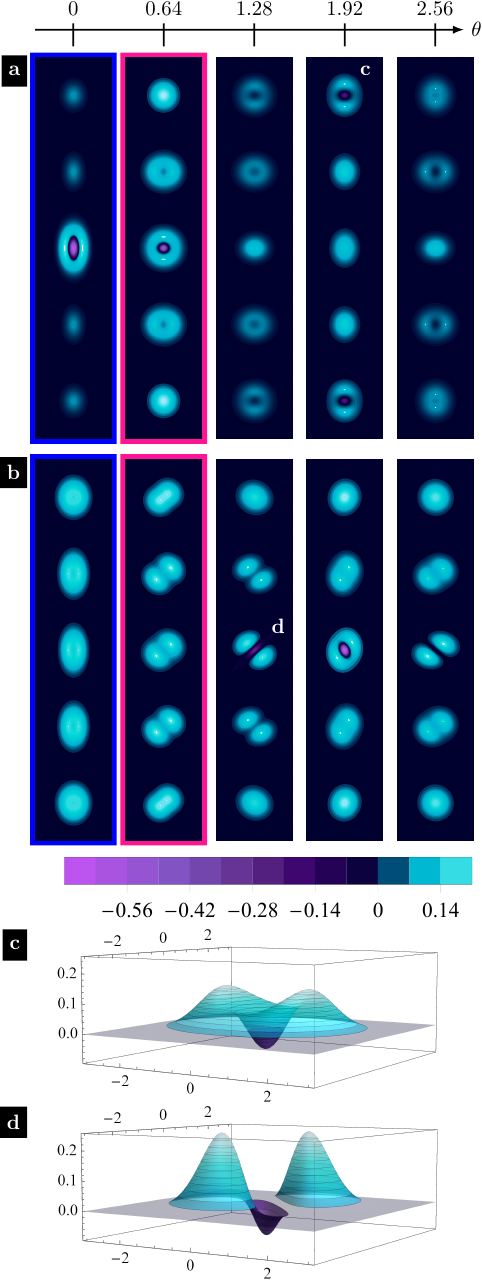}
 \caption{Evolution of the Wigner function for each mode in the pentamer when a single excitation is added to the initial state, which is the same product of single-mode squeezed states as in Section \ref{sc:squeezing}. The excitation is added in (a) the mode $a_3$ and (b) in a uniform superposition of the five modes. Blue and pink frames indicate the same stages as in Fig. \ref{fg:squeezing}. 3D plots in (c) and (d) show two illustrative examples of different evolved states, also indicated in the upper graphs.}
\label{fg:wigner}
\end{figure}

Upon this equivalence it is possible to keep track of how the characteristic function and the Wigner function of the system are transformed. Its definition, as derived in Ref. \cite{WalschaersPRA}, involves two key elements. The first one is the covariance matrix $V$, whose $i,j$-th
entry is the symmetrized covariance between quadrature operators $Q({\mathcal E}^s_{i})$ and  $Q({\mathcal E}^s_{j})$ evaluated in the basis $\varepsilon^s$. Following the same principle as in Eq. \eqref{eq:apsimp}, the entry $V_{ij}'$ after evolution is given by 
\begin{equation}\label{eq:covp}
  V_{ij}'=\frac{1}{2}\left\langle\Delta Q'({\mathcal E}^s_{i})\Delta Q'({\mathcal E}^s_{j})+\Delta Q'({\mathcal E}^s_{j})\Delta Q'({\mathcal E}^s_{i})\right\rangle,
\end{equation}
with
\begin{equation*}
\begin{split}
   Q'({\mathcal E}^s_{i})=
 &a\big( ({\rm Re}\,{\bf A}^{\rm T} \oplus {\rm Im}\,{\bf A}^{\rm T} ){\mathcal E}_i^s \big)\\
 &+a^\dagger\big( ({\rm Re}\, {\bf A}^{\rm T} \oplus -{\rm Im}\,{\bf A} ^{\rm T} ){\mathcal E}_i^s\big)\,.
 \end{split} 
 \end{equation*}
This matrix captures the influence of the Gaussian component of the initial state on the evolution of the system. Second, consider that a single excitation is added, the added or subtracted excitation in a mode is described by a vector $g$ in the symplectic basis. The additional correlations induced by this excitation are captured by the following matrix acting on the space $\mathbb{R}^{2N}$: 
\begin{equation}
 \mathcal A^{\pm}=2\frac{(V\pm 1)(P_g+P_{Jg})(V\pm  1)}{{\rm tr}\lbrace (V\pm 1)(P_g+P_{Jg})\rbrace}\,,
\end{equation}
where $P_g$ and $P_{Jg}$ are the projectors on vectors $g$ and $Jg$ respectively. Its evolution under $U$ is obtained by using Eq. \eqref{eq:covp} and transforming $g$ accordingly. Using these definitions (cf. Ref. \cite{WalschaersPRA}), we can express the Wigner function of the evolved state as
\begin{equation}\label{eq:Wadd}
 W^\pm({Y})=Z^\pm(Y)W_G(Y)
 \end{equation}
 with
 \begin{equation}\label{eq:Zpm}
 Z^\pm(Y)=\frac{1}{2}\left(Y^{\rm T}V^{-1}{\mathcal A}^\pm V^{\prime-1}Y-{\rm tr}(V^{\prime-1}{\mathcal A}^{\pm\prime})+2\right)\,,
\end{equation}
where $Y$ is a vector expressed in the base ${\mathcal E}^s$ which spans the quadrature space, and $W_G(Y)$ is the Wigner function of the non-displaced Gaussian state given by
\begin{equation}\label{eq:WGauss}
 W_G(Y)=\frac{1}{(2\pi)^N\sqrt{{\rm det}V'}}\exp\left(-\frac{1}{2}Y^{\rm T}V^{\prime-1}Y\right)\,.
\end{equation}

In Fig.~\ref{fg:wigner}, we present two illustrative examples showing the effect of adding a single excitation to the product of single-mode squeezed states studied in Sec. \ref{sc:squeezing} for the open pentamer ($N=5$). We observe that when the excitation is added to a single mode, it results in the expected negativity of the corresponding Wigner function around the center of the phase space. As the system evolves, dips appear in the Wigner functions of the adjacent modes, so even the functions of the edge modes exhibit a negative region at some point. On the other hand, when the single excitation is added as a uniform superposition in all the modes [Fig. \ref{fg:wigner} (b)], it does not lead to negative regions for any mode at the initial stage; interestingly, the Wigner functions distributions rotate through the evolution, with noticeable negativity regions appearing for the central mode. 

These examples show how highly non-Gaussian states arise during evolution, starting from products of quasi-Gaussian states. The symplectic space representation of field evolution allows studying different state superpositions added to a wide class of Gaussian states beyond our demonstrative results. 
Knowing the exact form of the Wigner function is crucial for applications such as computing the Cramer-Rao bound \cite{Pinel2012,Braun2014}6 for measuring phase shifts in a particular mode, thus providing a reliable estimation of the metrological improvement when using non-Gaussian input states.

\subsection{Effect of phase disorder on quantum correlations}\label{sc:corrs}

\begin{figure*}[!t]
 \includegraphics{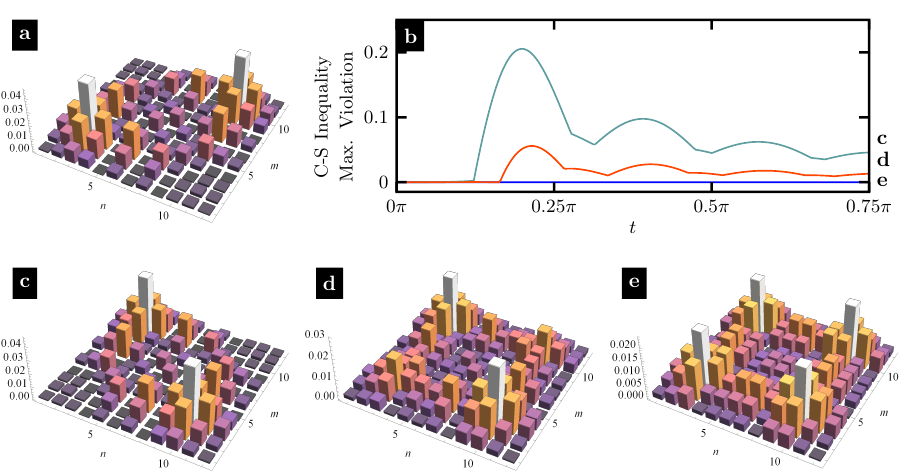}
 \caption{Two-particle correlations after evolution up to $t_f=3\pi/4$ from different initial states. (a) Initial state given by a product of single-particle states in adjacent middle modes, with real coupling. (b) Maximum violation of the Cauchy-Schwartz inequality. Corresponding cases are indicated by the letter next to the curves. (c) Evolution from a two-particle state in superposition over the adjacent center modes, $(\ket{2_{k}}+\ket{2_{\ell}})/\sqrt{2}$, for equal and real coefficients
 $\epsilon=0$. The same is shown in (d) for $\epsilon=0.55\pi$ and (e) for $\epsilon=\pi$. Results are averaged over 100 realizations.}\label{fg:corrs}
\end{figure*}

As a final application of our analytic results presented in Section \ref{sc:model}, we now study two-particle cross-correlations after propagation through an $N$-mer, highlighting the quantum properties of the propagating fields. In particular, the study of photon pairs propagating through waveguide arrays shows that the interference in the $N$-mode system gives rise to photon bunching resembling the Hong-Ou-Mandel effect \cite{Bromberg}. In the following, we study how this behavior is affected by decoherence arising from the random variation of the complex phases in the coupling coefficients. $N$-mode couplers exhibiting random amplitudes or \emph{weights} of the coupling constants are known to produce Anderson localization, extensively studied. However, the effect of disorder in the phases of the coupling coefficients remains, to our knowledge, unexplored.

Depending on the observable, two-particle states suffice to manifest the quantum nature of the system. A suitable measurement to reveal it is the two-particle correlation function $\Gamma_{mn}$, related to the probability of detecting a particle at each mode $m$ and $n$:
\begin{equation}\label{eq:corr}
 \Gamma_{mn}=\langle a^\dagger_m a^\dagger_n a_m a_n\rangle\,.
\end{equation} We are interested in computing the dynamics of the correlation $\Gamma_{mn}(t)$ for  a state $\ket{\psi_t}$. 

First, let us consider an initial two-photon state of the form $\ket{1_{k_0}}\ket{1_{k_0+1}}$ for two adjacent central modes $k_0$ and $k_0+1$, leading to 
\begin{equation}
 \ket{\psi_t}=U^{-1} \ket{1_{k_0}}\ket{1_{k_0+1}}= {a^\dagger_{k_0}}' {a^\dagger_{k_0+1}}'\ket{0}.
\end{equation}


Substituting in Eq.~\eqref{eq:corr} and using the result of Eq.~\eqref{eq:Amn} we obtain
\begin{equation}\label{eq:Gammaprod}
\begin{split}
 \Gamma_{mn}(t)=&|A^{(N)}_{k_0,m}A^{(N)}_{k_0+1,n}+A^{(N)}_{k_0,n}A^{(N)}_{k_0+1,m}|^2\\
 =&\frac{1}{2}|\beta^{(N)}_{k_0,m}\beta^{(N)}_{k_0+1,n}|^2+\frac{1}{2}|\beta^{(N)}_{k_0,n}\beta^{(N)}_{k_0+1,m}|^2\\
 &+\beta^{(N)}_{k_0,m}\beta^{(N)}_{k_0+1,n}\beta^{(N)\ast}_{k_0,n} \beta^{(N)\ast}_{k_0+1,m}\\ &\times \exp\left(-i[\phi_{k_0,m}+\phi_{k_0+1,n}-\phi_{k_0,n}-\phi_{k_0+1,m}]\right)\\&+{\rm h.c.}\,,
\end{split}
\end{equation}
where we have used the notation introduced in Section~\ref{sc:tb}
in order to make explicit the dependence of $\Gamma_{mn}$ on phases $\delta_j$, which is then contained in the exponential term in Eq.~\eqref{eq:Gammaprod}. By noticing that $ \phi_{j,m}-\phi_{j,n}=\phi_{n,m}$ and $\phi_{m,n}=-\phi_{n,m}$, we see that the argument in the exponential cancels, so $\Gamma_{mn}$ is independent of phases $\delta_j$. Accordingly, decoherence does not affect the dynamics of a product state since it does not exhibit coherence in the form of interference between two modes. Therefore, for any complex coupling phases, the evolution of $\ket{1_{k_0}}\ket{1_{k_0+1}}$ leads to the well-known bunching effect obtained from two-particle product states \cite{Peruzzo,Bromberg}, where both particles are most likely to end in the same mode, as shown in Fig.~\ref{fg:corrs}(a). Note that the dynamics of the product state in the $N$-mer saturates the Cauchy-Schwartz inequality $\Gamma_{mn}<\sqrt{\Gamma_{mm}\Gamma_{nn}}$ but never violates it, as shown in Fig.~\ref{fg:corrs}(b).

Let us now consider that the system is initially in a superposition of two-photon states in modes $k$ and $\ell$, i.e.,
\begin{equation}\label{eq:sup}
 \ket{\psi_0}=\frac{1}{\sqrt{2}}\left(\ket{2_{k}}+\ket{2_{\ell}}\right)\,.
\end{equation}

The state after time $t$ is given by
\begin{equation}\label{eq:supt}
\begin{split}
 \ket{\psi_t}=\frac{1}{\sqrt{2}}\left((a'_{k})^{\dagger 2}+(a'_{\ell})^{\dagger 2}\right)\ket{0}\,.
 \end{split}
\end{equation}

Replacing Eqs.~\eqref{eq:Amn} and \eqref{eq:supt} in Eq.~\eqref{eq:corr} we obtain
\begin{equation}\label{eq:Gammasup}
\begin{split}
 \Gamma_{mn}(t)=&|A^{(N)}_{k,m}A^{(N)}_{k,n}+A^{(N)}_{\ell,m}A^{(N) }_{\ell,n}|^2\\
 =&|\beta^{(N)}_{k,m}\beta^{(N)}_{\ell,n}|^2+|\beta^{(N)}_{k,n}\beta^{(N)}_{\ell,m}|^2\\
 &+\beta^{(N)}_{k,m}\beta^{(N)}_{\ell,n}\beta^{(N)\ast }_{k,n} \beta^{(N)\ast}_{\ell,m}\\ &\times 2\cos(2\phi_{k,\ell})+{\rm h.c.}\,
\end{split}
\end{equation}

Unlike the product state, the initial superposition in Eq.~\eqref{eq:sup} leads to a final state depending on the phases $\delta_j$ of the coupling coefficients. If $\ket{\psi_0}$ is composed of only two adjacent modes $k_0$ and $k_0+1$, the cosine in Eq.~\eqref{eq:Gammasup} simplifies to $\cos(2\delta_{k_0})$. In this case, a disorder in the coefficient phases reduces to $\delta_{k_0}$ varying randomly with a normal distribution of variance $\epsilon$ centered around $\delta_{k_0}=0$. The width $\epsilon$ defines the degree of disorder. The average can be computed by direct integration leading to
\begin{equation}\label{eq:Gammasupav}
\begin{split}
 \Gamma_{mn}(t)=&|\beta^{(N)}_{k_0,m}\beta^{(N)}_{k_0+1,n}|^2+|\beta^{(N)}_{k_0,n}\beta^{(N)}_{k_0+1,m}|^2\\
 &-\exp(-2\epsilon^2)\beta^{(N)}_{k_0,m}\beta^{(N)}_{k_0+1,n}\beta^{(N)\ast}_{k_0,n} \beta^{(N)\ast}_{k_0+1,m}\\
 &+{\rm h.c.}\,
\end{split}
\end{equation}

In Fig.~\ref{fg:corrs}(c)-(d), we present the two-photon correlation function at time $t_f=3\pi/4$ for different decoherence degrees (normal distribution of width $\epsilon$). For $\epsilon=0$ (no decoherence, identical and real coupling coefficients), a superposition of two-photon states in two adjacent modes $k=k_0$ and $\ell=k_0+1$ leads to the expected behavior for the correlations shown in Fig.~\ref{fg:corrs}(c), with each of the photons ending in distant modes, instead of the bunching obtained if both particles start in a product state. This effect is due to the interference expressed by the last term in Eq.~\eqref{eq:Gammasupav}. As we raise $\epsilon$ to introduce disorder in the phases $\delta_j$, we observe how this behavior is gradually suppressed, and the output correlation becomes similar to the one obtained from a macroscopic or classical state. Again, this can be understood from Eq.~\eqref{eq:Gammasupav}, since the term accounting for quantum interference decays as $\exp(-2\epsilon^2)$. In fact, for $\epsilon=\pi$, Fig.~\ref{fg:corrs}(e), the correlation matrix exhibits a four-peak pattern similar to the one obtained from an initial state $\ket{2}_{k_0}$, which is a simple product of two single-photon distributions without coherence \cite{Bromberg}. Interestingly, this is also reflected by a violation of the Cauchy-Schwartz inequality. Fig.~\ref{fg:corrs}(b) shows the maximum violation as a function of $t$ for the scenarios under consideration. We observe that when the system evolves from an initial superposition, the inequality is violated as far as the decoherence degree $\epsilon$ is low; as this increases and the interference is suppressed, the maximum violation gradually decays until the system just saturates the Cauchy-Schwartz inequality for large $\epsilon$. The violation of the inequality serves as a criterion to test whether the quantum properties of the state are fragile to decoherence, in this case, induced by random phases in the coupling coefficient.

The analytic solution of the TB model with complex coefficients allowed us to describe a previously unexplored decoherence mechanism, where quantum correlations wash out after propagation through an $N$-mer with random coupling phases. More generally, the description allows for simple expressions describing higher-order spatial correlation between modes, an otherwise challenging computational task.

\section{Benchmark}

\begin{figure}[b!]
\centering
 \includegraphics[width=.4\textwidth]{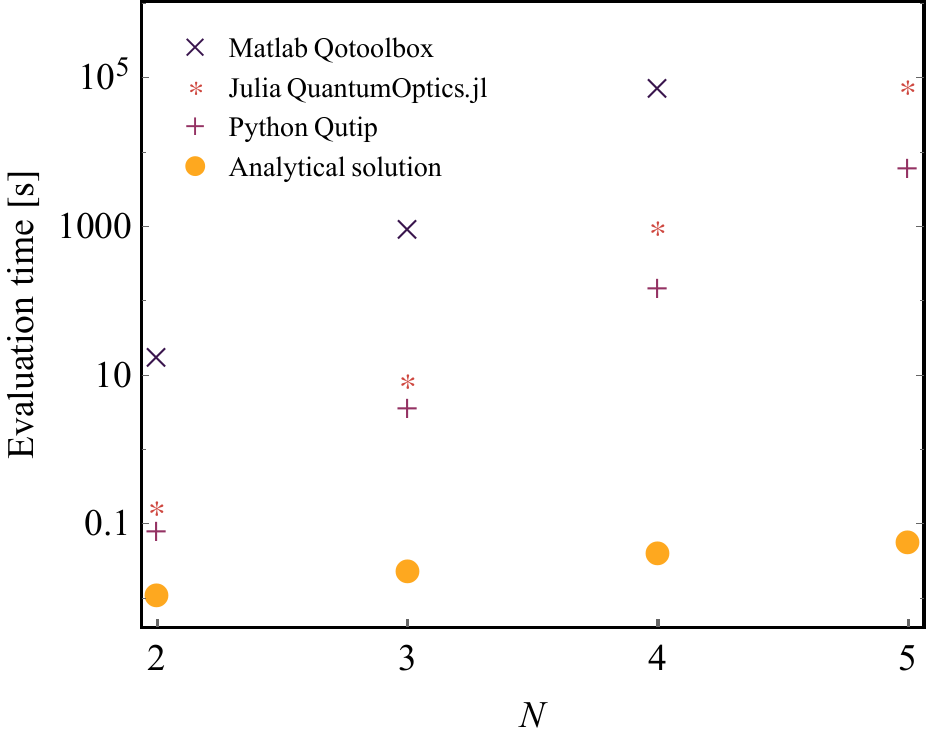} \caption{Comparison of the average time required to compute the evolution of squeezed initial states in the $N$-mer using numerical integration in different languages and the analytical results of Sec. \ref{sc:A}. 
 }\label{fg:benchmarks}
\end{figure}


The standard procedure for solving the evolution of quantum systems often involves employing a basis function set spanning the Hilbert space, diagonalizing the Hamiltonian matrix, and using tensor products and partial traces to derive the desired expectation values. This procedure is challenging unless the dimension of the Hilbert space is small \cite{qutip}. The computational complexity grows exponentially with the number of modes \cite{Bromberg}, emphasizing the need for efficient computational tools.

We now compare solving the $N$-mer evolution with our analytical expressions versus numerical solvers of the Schr\"odinger equation. Figure \ref{fg:benchmarks} compares the computation time required to calculate the evolution from initial products of squeezed vacuum states with real squeezing parameters (similar to the initial states used in Sec. \ref{sc:squeezing}) 
using the respective toolboxes for Matlab, Julia, Python, and our analytical results. 
Solutions in these languages were computed with a maximum particle number equal to 40, which leads to a respective Hilbert space of size 1600 for $N=2$, 64\,000 for $N=3$, 2\,560\,000 for $N=4$ and 102\,400\,000 for $N=5$. 

All calculations were carried out on a laptop computer with an 8-core Intel i7 processor operating at a clock speed of 2.80 GHz, and 16 GB of DDR4 RAM. The numerical values of the analytical expression were obtained by implementing the respective functions in C++. The results showcase a substantial advantage in average computation time when utilizing analytical expressions, reaching up to six orders of magnitude difference for the pentamer case. We omit tests for larger systems due to RAM limitations, which introduce significant inconsistencies in computation time for numerical approaches, where resource allocation becomes crucial. In fact, for $N > 6$, the equations could not be solved numerically, even on a computer with 128 GB RAM.

\section{Summary and Outlook}\label{sc:cons}

Initially proposed to calculate band structures in solids, the tight-binding model describes several analogous systems, such as optical lattices with cold atoms, light propagating in phononic crystals, surface waves, topological insulators, and quantum random walks. Despite its conceptual simplicity and broad applicability, many cases lack analytical solutions. We find exact analytical solutions for the spatial evolution of arbitrary quantum states in one-dimensional tight-binding models with complex coupling constants of equal amplitudes and arbitrary phases. Our results have profound implications for understanding transport dynamics and quantum correlations in real space, enabling us to describe these phenomena within an analytical framework. 

Our analytic solutions provide a considerable computational advantage over standard numerical solvers regarding efficiency and accuracy. In particular, we compared performances when solving the Schr\"odinger equation to compute spatial propagation and correlations in large systems involving many particles for non-trivial initial conditions, such as products of single-mode squeezed states. 

To demonstrate the generality and practicality of our analytic results, we have applied them to the study of quantum light propagation in multimode linear couplers. We have showcased the generalization of multimode squeezing formation from single-mode squeezed states and efficiently computed the dynamics of the Wigner function of single excitation added (subtracted) to (from) a displaced many-modes Gaussian state, revealing the emergence and propagation of highly non-Gaussian states. Finally, we also studied the evolution of two-particle cross-correlations, highlighting the propagation of quantum properties of the fields. In particular, we analytically showed that decoherence arising from phase disorder, that is, random variation of the complex phases in the coupling coefficients, leads to a quantum-to-classical transition in the two-particle correlation function. 

Although our main emphasis is on a mathematical method to obtain analytical solutions of the tight-binding model, we also provide new physical insight in its application to interesting examples. However, we hope that our results may prove valuable in the study of other physical systems described by the tight-binding model, such as, for instance, other geometric configurations. In this regard, analytic results such as ours could have an impact extending as broad as the tight-binding model itself.\\

{\it Acknowledgments.---} 
This work was supported in part by ANID-PAI grants N$^{\circ}$ 77180003 and N$^{\circ}$ 77190033, FONDECYT grants N$^{\circ}$ 1230897, XXX, and N$^{\circ}$ 11200192, ANID Doctoral fellowship grant N$^{\circ}$ 21192248 and ANID - Millenium Science Inititive Program - ICN17\_012. A.D. was supported by FONDECYT Grants No. 1231940 and No. 1230586.

\setcounter{equation}{0}
\appto\appendix{\counterwithin{equation}{section}}
\appendix
\section{Coupling matrix}
\label{Appendix:coupling_matrix}

The transformed mode $a^{\prime}_m$ according to Eq~\eqref{eq:expBCHb} is given by
\begin{equation}\label{eq_appendix:App_BCH}
    \begin{split}
    a^{\prime}_m &= \sum_{n=0}^\infty\frac{1}{n!}\left(\frac{it}{\hbar }\right)^n\left[H,a_m\right]_n,
    \end{split}
\end{equation}
where
\begin{equation}\label{eq:apenH}
    \begin{split}
        \left[ H, a_m\right]_{n} &= \sum_{j=1}^N \Lambda_j^{(n)} a_j.
    \end{split}
\end{equation}
Now we calculate the $(n+1)$-th commutator between $H$ and $a_m$, obtaining
\begin{equation}
    \begin{split}
        \left[ H, a_m\right]_{n+1} &= \left[ H, \sum_{l=1}^N \Lambda_l^{(n)} a_l\right].
    \end{split}
\end{equation}
Using the general Hamiltonian Eq.~\eqref{eq:H} in the equation above the commutator becomes
\begin{equation}
    \begin{split}
        \left[ H, a_m\right]_{n+1} &= -\sum_{l=1}^N \hbar\Lambda_l^{(n)}  \left( \sum_{\langle j,k\rangle}\left( C_{j,k}\delta_{lj}a_k+C_{j,k}^\ast\delta_{lk}a_j\right) \right).
    \end{split}
\end{equation}
Since the sum is constrained to modes $j$ and $k$ connected to mode $l$, we obtain 
\begin{equation}\label{eq:App_commutation_n+1_1}
    \begin{split}
        \left[ H, a_m\right]_{n+1} &= -\sum_{l=1}^N \hbar\Lambda_l^{(n)}  \left( \sum_{ j(l),k(l)}\left(\vphantom{C_{j(l),l}^\ast}\right.\right. C_{l,k(l)} a_{k(l)}\\
        &+\left.\left.C_{j(l),l}^\ast a_{j(l)}\right) \vphantom{\sum_{ j(l),k(l)}}\right).
    \end{split}
\end{equation}
Using the fact that $C_{i,j}^{\ast}=C_{j,i}$ and by exchanging the summation over equation \eqref{eq:App_commutation_n+1_1} we find for the $(n+1)$-th commutator
\begin{equation}\label{eq:App_commutation_n+1_2}
    \begin{split}
        \left[ H, a_m\right]_{n+1} &= \sum_{j=1}^N\left(-\hbar\sum_{l(j)}\Lambda^{(n)}_{l(j)} C_{l(j),j}\right)a_j.
    \end{split}
\end{equation}
From \eqref{eq:App_commutation_n+1_2} we can see that the $\Lambda^{(n+1)}_j$ coefficients are recursively related to the $\Lambda^{(n)}_{l(j)}$ coefficients by the iterative rule
\begin{equation}
    \begin{split}
        \Lambda^{(n+1)}_j &= -\hbar\sum_{l(j)} C_{l(j),j}\Lambda^{(n)}_{l(j)}.
    \end{split}
\end{equation}
This can be cast as a matrix equation
\begin{equation}
\vec \Lambda_{n+1} = -\hbar\mathcal{C}^\intercal\vec\Lambda_n,
\end{equation}
where
\begin{equation}\label{eq:App_recursion}
    \begin{split}
        \mathcal{C} =
        \begin{pmatrix}
          0 & C_{1,2} & \cdots & C_{1,N}\\
          C^\ast_{1,2} & 0 &\cdots & C_{2,N}\\
          \vdots & \vdots & \ddots & \vdots\\
          C^\ast_{1,N} & C^\ast_{2,N} &\cdots & 0
        \end{pmatrix}\,,\ 
        \vec \Lambda_{n}=
        \begin{pmatrix}
          \Lambda^{(n)}_1\\
          \Lambda^{(n)}_2\\
          \vdots\\
          \Lambda^{(n)}_N
        \end{pmatrix}
    \end{split}
\end{equation}


The recursive relation between the $\Lambda$ coefficients in equation \eqref{eq:App_recursion}  guides us to apply $\mathcal{C}^\intercal$ $n$ times over an initial condition $\vec\Lambda_0$ (which depends on the transformed mode $a_m$). By doing this operation the recursive commutator $[H,a_m]_n$ results in
\begin{equation}\label{eq:App_commutator_n}
    \begin{split}
    \left[H,a_m\right]_n &= \sum_{j=1}^{N}\hat e_j^\intercal \left(-\hbar\mathcal{C}^\intercal\right)^n\hat e_m a_j.
    \end{split}
\end{equation}
Inserting \eqref{eq:App_commutator_n} in \eqref{eq_appendix:App_BCH} we obtain for the transformed modes 
\begin{equation}\label{a_propagado}
    \begin{split}
        a^{\prime}_m &= \sum_{n=1}^{\infty}\frac{1}{n!}\left(\frac{it}{\hbar}\right)^n\left(\sum_{j=1}^N \hat e_j^\intercal \left(-\hbar\mathcal{C}^\intercal\right)^n\hat e_m a_j\right),\\
    a^{\prime}_m &=\sum_{j=1}^N \hat e_j^\intercal \exp{\Big(-it\mathcal{C}^\intercal\Big)}\hat e_m a_j.
    \end{split}
\end{equation}
Thereby, transformed modes $a^{\prime}$ are related to the modes $a_j$ through the matrix relation
\begin{equation}
    \begin{pmatrix}
      a^{\prime}_1\\
      a^{\prime}_2\\
      \vdots\\
      a^{\prime}_N
    \end{pmatrix}
    =
    \exp{\left(-it\mathcal{C}\right)}
    \begin{pmatrix}
      a_1\\
      a_2\\
      \vdots\\
      a_N
    \end{pmatrix}.
\end{equation}


\section{Main results}
\label{Appendix:Main results}
In this section we summarize the main results obtained when evaluating the matrix
\begin{equation}
A=\exp{\left(-it\mathcal{C}\right)},
\end{equation}
where $\mathcal{C}$ is the coupling coefficient matrix for closed and open arrays. A detailed derivation of the coefficients of $A$ can be found in Appendix \ref{Appendix:Derivation}
\subsection{a. Closed array}
In this case, the matrix $A$ turns out to be circulant. Thereby, the $n$-th row has the same entries as the first row but cycled by $n$ positions. The first-row coefficients of matrix $A$ for a $N$-mer with first-neighbor coupling constants given by $C_j=C\exp{(i\delta)}$ with $C$ and $\delta$ arbitrary but contant are given by the expression
\begin{equation}
     \begin{split}
         \overline{A}_{1,n}^{(N)} =
         \begin{cases}
         \alpha_{N}^{(1)}\; &N\ \text{odd},\ n=1\\
         \alpha_{N,n}^{(2)}\; &N\ \text{odd},\ n\leq(N+1)/2\\
         \mathcal{P}_c\alpha_{N,N-n+2}^{(2)} \; &N\ \text{odd},\ n>(N+1)/2\\
         \alpha_{N}^{(0)}\; &N\ \text{even},\ n=1\\
         \alpha_{N,n}^{(2)}\; &N\ \text{even},\ n\leq N/2\\
         \alpha_{N,n}^{(3)} \; &N\ \text{even},\ n=\frac{N}{2}+1,\ n \text{ odd}\\
         \alpha_{N,n}^{(4)} \; &N\ \text{even},\ n=\frac{N}{2}+1,\ n \text{ even}\\
         \alpha_{N,N-n+2}^{(2)\ast}\;  &N\ \text{even},\ n>\frac{N}{2}+1,\ n\text{ odd}\\
         -\alpha_{N,N-n+2}^{(2)\ast}\; &N\ \text{even},\ n>\frac{N}{2}+1,\ n\text{ even}
         \end{cases}
     \end{split}
 \end{equation}
where the $\alpha_{N,n}$ functions are given by
 \begin{align}
    \nonumber \alpha_{N}^{(0)} &= -J_0(2Ct)+2\sum_{l=0}^{\infty}
         \cos{(lN(\delta-\pi/2))}J_{lN}(2Ct)\,, \\  
    \nonumber \alpha_{N}^{(1)} &= -J_0(2Ct)\\
    \nonumber &+2\sum_{l=0}^{\infty}\Big(\cos{(2lN(\delta-\pi/2))}J_{2lN}(2Ct)\\
    \nonumber &+i\sin{((2l+1)N(\delta-\pi/2))}J_{(2l+1)N}(2Ct)\Big)\,,\\
    \nonumber \alpha_{N,n}^{(2)} &= \sum_{l=1}^{\infty}\sigma_{N,n,l}\exp{\left\{(-1)^{l+1}b_l(N,n)(\delta-\pi/2) i\right\}}\\
    \nonumber &\times J_{b_l}(2Ct)\,,\\
    \nonumber \alpha_{N,n}^{(3)} &= 2\sum_{l=0}^{\infty} \cos{\left(\frac{N}{2}(2l+1)(\delta-\pi/2)\right)}J_{N(2l+1)/2}(2Ct)\,, \\
    \alpha_{N,n}^{(4)} &= 2i\sum_{l=0}^{\infty} \sin{\left(\frac{N}{2}(2l+1)(\delta-\pi/2)\right)}J_{N(2l+1)/2}(2Ct)\,,
 \end{align}
with a parity conjugation operator
\begin{equation}
    \mathcal{P}_c e^{i n\delta}=
    \begin{cases}
    -e^{-i n\delta}\quad n \text{ odd}\\
    \quad\! e^{-i n\delta} \quad n \text{ even},
    \end{cases}
\end{equation}
the Bessel functions $J_k(2Ct)$ of first kind and the special sign function $\sigma_{N,n,l}$ is given by
and 
\begin{equation}
\begin{split}
    \sigma_{N,n,l} &=
    \begin{cases}
    \sigma_{1}(b_l(N,n))\quad  N \text{ odd } ,\ n \text{ even} \\
    \sigma_{2}(b_l(N,n)) \quad  N \text{ odd },\ n \text{ odd}\\
    (-1)^{l+1} \qquad\quad\! N \text{ even } ,\ n \text{ even} \\
    \quad\, 1 \quad\qquad\qquad\!\!\! N \text{ even } ,\ n \text{ odd}
    \end{cases},\\
    \sigma_{1}(b_l(N,n)) &= 
        \begin{cases}
            -1\quad \ b_l(N,n)=2N-n+1 \ {\rm mod}(2N)\\
            \;\;\; 1\quad \text{Otherwise}
        \end{cases}\\
        \sigma_{2}(b_l(N,n)) &= 
        \begin{cases},
            -1\quad \ b_l(N,n)=N-n+1 \ {\rm mod}(N)\\
            \;\;\; 1\quad \text{Otherwise}
        \end{cases},
         \end{split}
\end{equation}
where indexes $b_l(N,n)$ are defined by the set 
\begin{equation}
    b_l(N,n) = \{k\in\mathbb{N}\ |\ k= n-1 \ \vee\  N-n+1\ {\rm mod}(N) \}.
\end{equation}

\subsection{b. Open array}
In this case, The coefficients of matrix A for a $N$-mer with first-neighbor coupling constants given by $C_j=C\exp{(i\delta_j)}$ with $C$ and $\delta_j$ arbitrary are given by the expression
\begin{equation}
 \begin{split}
 A_{m,n}^{(N)}= \frac{2}{N+1}\sum_{k=1}^N \exp\left(-2i\cos\left[\frac{k\pi}{N+1}\right]C t\right)S_{mnk}\,,
 \end{split}
\end{equation}
where
\begin{equation}
\begin{split}
  S_{mnk}=&\exp\left(-i\left[\frac{\pi}{2}(m-n)+\Delta_{m-1}-\Delta_{n-1}\right]\right)\\
 &\times \sin\left(\frac{m k\pi}{N+1}\right)\sin\left(\frac{n k\pi}{N+1}\right)
\end{split}
\end{equation}
and 
\begin{equation}
\Delta_n=\delta_1+\delta_2+\cdots+\delta_n.
\end{equation}

\section{Derivation}
\label{Appendix:Derivation}

\subsection*{a. Mathematical context: overview}
We consider a lattice formed by $N$ nearest-neighbor evanescently coupled bosonic single-mode described by the quantum Hamiltonian
\begin{equation}\label{eq_appendix:H_open}
 H=\hbar \sum\limits_{j=1}^{N-1}\left(C_j a_j^\dagger a_{j+1}+C_j^\ast a_{j+1}^\dagger a_j\right)\,,
\end{equation}
where $a_j$ ($a_j^\dagger$) is the annihilation (creation) operator of an excitation in the $j$-th mode and $C_j=C\exp(i\delta_j)$ is the complex coupling coefficient between the $j$-th and the $(j+1)$-th modes, with $C$ and $\delta_i$ real numbers.
It is worth noting that the standard model used for waveguide arrays or Bose-Einstein condensates in periodical lattices, which considers real coupling constants, is obtained by taking $\delta_j\equiv0$, i.e. $C_j\equiv C$.

Closed analytic solutions for the elements of the transformation matrix $A^{(N)}$ Eq.~\eqref{eq:expCoupling} were obtained for $N=2$ \cite{jmo} and $N=3$ \cite{nosotros}. For $N=4$, the transformation element that accompanies $a_2$ in the evolution of $a_1$ (that is, $a_1^{\prime}$), namely $A_{1,2}^{(4)}\,$, can be obtained in terms of the series expansion of Eq.~\eqref{eq_appendix:App_BCH} resulting in
\begin{equation}\label{eq_appendix:A12_open}
    \begin{split}
        A_{1,2}^{(4)} &= -ie^{i\delta_1}\left((Ct)-\frac{2}{3!}(Ct)^3+\frac{5}{5!}(Ct)^5-\frac{13}{7!}(Ct)^7\right.\\
    &\left.+\frac{34}{9!}(Ct)^9-\frac{89}{11!}(Ct)^{11}+\frac{233}{13!}(Ct)^{13}-\cdots \right)\,.
    \end{split}
\end{equation}
The coefficients in Eq.~\eqref{eq_appendix:A12_open} exhibit an interesting property, they are the odd terms of the Fibonacci succession, so the series can be rewritten as
\begin{equation}\label{eq_appendix:A12_Sj}
  A_{1,2}^{(4)}=-ie^{i\delta_1}\sum\limits_{j=0}^\infty \frac{(-1)^j}{(2j+1)!} S_{2j+1} \left(C t\right)^{2j+1}.
\end{equation}
By means of Binet's Fibonacci Number Formula \cite{spickerman1982binet}, the $j$-th term of the succession is expressed as
\begin{equation}\label{eq_appendix:S_j}
    S_j = \frac{1}{\sqrt{5}}\left(\left(\frac{1+\sqrt{5}}{2}\right)^j-\left(\frac{1-\sqrt{5}}{2}\right)^j\right)\,,
\end{equation}
Using Eq.~\eqref{eq_appendix:S_j} in Eq.~\eqref{eq_appendix:A12_Sj} we finally obtain an analytic expression for $A_{1,2}^{(4)}$
\begin{equation}\label{eq_appendix:A4_12}
    A_{1,2}^{(4)} = -\frac{ie^{i\delta_1}}{\sqrt{5}}\left(\sin{\left(\frac{1+\sqrt{5}}{2}Ct\right)}+\sin{\left(\frac{\sqrt{5}-1}{2}Ct\right)}\right)\,.
\end{equation}
The previous calculation of $A_{1,2}^{(4)}$ is essential for acquiring a comprehensive understanding of the procedure to achieve general analytic expressions of $A^{(N)}$. After the expansion of Eq.~\eqref{eq_appendix:App_BCH}
\begin{equation}\label{eq_appendix:definicionBeta}
    A_{m,n}^{(N)}\propto \beta^{(N)}_{mn}=\sum_{k=0}^\infty F_{mnk} (Ct)^k\,,
\end{equation}
the main task is to identify the behavior of the series $\beta^{(N)}_{m,n}$, that is, the rule behind the terms $F_{mnk}$. Therefore, efficient computation of the amplitudes of the output modes requires knowing the convergence of the series expansion $\beta^{(N)}_{m,n}$ from the BCH formula for any $N$. This series expansion is related to several interesting problems in mathematics, as we will now discuss.


For $N\geq 5$, the coefficients in the series expansions are associated with less known successions. Table~\ref{tb:seqs} presents the sequences appearing in the expansions of $\beta_{1,1}^{(N)}$ and $\beta_{N,N}^{(N)}$ for $N\leq8$, identified by their respective OEIS codes \cite{oeis}. For $N>8$ we cannot unequivocally identify the sequences with the current OEIS database. We notice that the sequence expanding the entries of the evolution matrix $A^{(N)}$ of an $N$-mode system differs from those of an $N-1$-mode system at the $N$-th place (highlighted with bold font in Table~\ref{tb:seqs}). The sequences clearly show that the effect of adding a mode to a $(N-1)$-mode system manifests at the $N$-th order in the series expansion for the output operator. 
Such sequences can be found recursively for arbitrary $N>2$ using their \emph{generating function} $G_N(x)$ defined by $G_N(x)=1/(1-xG_{N-1}(x))$, with $G_2(x)=1/(1-x)$, show in Table \ref{tb:seqs}.
\begin{table}[!ht]
\begin{tabular}{|c|c|c|c|} \hline
  $N$ & Sequence & OEIS & $G_N(x)$ \\ \hline
  2  & 1,{\bf 1},1,1,1,1,1,\ldots & A000012 & $\frac{1}{1-x}$ \\ \hline
  3  &1,1,{\bf 2},4,8,16,32,\ldots &A011782 & $\frac{1-x}{1-2x}$ \\\hline
  4  &1,1,2,{\bf 5},13,34,89,233,\ldots &A001519 & $\frac{1-2x}{1-3x+x^2}$ \\\hline
  5  &1,1,2,5,{\bf 14},41,122,365,\ldots &A124302 &$\frac{1-3x+x^2}{1-4x+3x^2}$ \\\hline
  6  &1,1,2,5,14,{\bf 42},131,417,\ldots &A080937 &$\frac{1-4x+3x^2}{1-5x+6x^{2}-x^{3}}$ \\\hline
  7 &1,1,2,5,14,42,{\bf 132},428,\ldots &A024175 &$\frac{1-5x+6x^{2}-x^{3}}{1-6x+10x^{2}-4x^{3}}$  \\\hline
  8 &1,1,2,5,14,42,132,{\bf 429},\ldots &A080938 &$\frac{1-6x+10x^{2}-4x^{3}}{1-7x+15x^{2}-10x^{3}+x^{4}}$ \\\hline
 \end{tabular}

 \caption{Sequences in the coefficients $\beta_{1,1}$ and $\beta_{N,N}$ for different $N$. We show the respective OEIS code and generating function.}\label{tb:seqs}
\end{table}

Finding the generating function $G_N(x)$ for all the matrix coefficients $\beta_{i,j}$ is more complicated, although it is still a feasible approach to obtain the respective sequences. However, the effect of having arbitrary phases $\delta_i$ in the coupling coefficients requires a specific definition of $G_{N}$ for each coefficient, so we must take a different approach for the present case.


\subsection*{b. General relations}

For the open array (Fig. \ref{fg:modelo} (b)) the complex entries of the transformation matrix $A^{(N)}$ can be generally expressed as
\begin{equation}\label{eq_appendix:A_mn}
 A^{(N)}_{m,n}=\exp(-i\phi_{m,n})\beta^{(N)}_{m,n}.
\end{equation}
An evaluation of the recursive terms in Eq.~(\ref{eq:apenH}) leads to
\begin{equation}\label{eq_appendix:definicionPhi}
 \phi_{m,n}=\Delta_{m-1}-\Delta_{n-1}\,,
\end{equation}
with $\Delta_n=\delta_1+\delta_2+\cdots+\delta_n$. All the information about the complex phases of the respective coupling coefficients is captured by $\phi_{m,n}$.
Such a simple expression for the phase $\phi_{m,n}$ is a unique feature of the open one-dimensional model. As discussed in Ref. \cite{lu2016}, this and other particular geometries allow the unitary operation with complex coupling to be decomposed in three stages: evolution with real-valued coupling preceded and succeeded by individual phase shifts on each mode. In this geometry, the asymmetry introduced by the complex coupling coefficients is not expected to affect the transition probabilities from one mode to the other, but only the transition amplitudes. 

For the closed $N$-mer (Fig.~\ref{fg:modelo}(c)) a factorization of the phase $\phi_{mn}$ analog to Eq.~\eqref{eq_appendix:definicionPhi} could not be found since the entries of $A^{(N)}$ behaves differently from the open array. Considering equal phases $\delta_j=\delta$, the evaluation of Eq.~\eqref{eq:apenH} leads to expressions of the form
\begin{equation}\label{eq_appendix:A_1n_form}
    \begin{split}
        \overline{A}_{1,n}^{(N)} &= \sum_{l=1}^{\infty}\sigma_{N,n,l}\exp{\left\{(-1)^{l+1}b_l(N,n)\delta i\right\}}\beta_{n b_l}^{(N)}
    \end{split}
\end{equation}
where $\sigma_{N,k,l}$ is sign function; $b_l(N,n)$ are a set of integers numbers and $\beta_{nb_l}^{(N)}$ follows the definition in Eq.~\eqref{eq_appendix:definicionBeta}.\\
\indent As can be seen in Eq.~\eqref{eq_appendix:A_1n_form}, the phase of the complex coupling coefficients are embedded in every term of the expansions. Although there is no factorization of the phases like Eq.~\eqref{eq_appendix:definicionPhi} for the close $N$-mer, what only changes with arbitrary phases is that $\delta$ become a functions of all the phases $\delta=\delta(\delta_1,\delta_2,\cdots,\delta_N)$.

\subsection*{c. Successive method}
\label{Appendix:Successive_method}
 As was pointed out in the case of $A_{1,2}^{(4)}$ (Eq.~\eqref{eq_appendix:A4_12}), $F_{mnk}$ is linked to integer sequences. If the nature of these numerical successions is recursive, we can write them as
\begin{align}\label{eq_appenddix:Sn}
 S_{n+r}^{} &= Q_1 S_{n+r-1}^{}+Q_2S_{n+r-2}^{}+\cdots+ Q_r S_n^{}\,,
 \end{align}
where $r$ is the recursive order and $Q_r$ integer numbers (by instance, for the Fibonacci sequence $r=2$ and $Q_1=Q_2=1$). Eq.~\eqref{eq_appenddix:Sn} can be cast into matrix form
\begin{equation}\label{eq:recur}
{\bf S}_{j}^{(N)}=R^{(N)}\,{\bf S}_{j-1}^{(N)},
\end{equation}
being $R^{(N)}$ the recursion matrix and ${\bf S}^{(N)}_j$ a vector composed of $r$ successive terms in the sequence, i.e.
\begin{align}
    {\bf S}_{j}^{(N)} &=\begin{pmatrix}
    S_{j+r-1}^{} \\
    S_{j+r-2}^{} \\
    \vdots \\
    S_{j}^{}
    \end{pmatrix}\,,\\
    R^{(N)} &= \begin{pmatrix}
    Q_1 & Q_2 & \cdots &Q_{r-1} & Q_r \\
     1  & 0   & \cdots & 0 & 0   \\
     0  & 1   & \cdots & 0 & 0   \\
     \vdots&\vdots&\ddots& \vdots &\vdots\\
     0  & 0   &\cdots & 1 & 0
    \end{pmatrix}
\end{align}
The superscript $(N)$ is written to indicate the relationship to the transformation matrix $A^{(N)}$. The Eq.~\eqref{eq:recur} can be rewritten by decomposing the initial condition ${\bf S}_0^{(N)}$ on the eigenvectors $\lbrace{\bm \varphi}_j\rbrace$ of the recursion matrix $R^{(N)}$. This leads to the following expression for the $j$-th term in the sequence
\begin{equation}\label{eq:Snb}
 S_j^{(N)}=\sum\limits_{l=1}^r\eta_l^{}\varphi_l^j\,,
\end{equation}
where $\varphi_l$ is the $l$-th eigenvalue of $R^{(N)}$. In Eq.~\eqref{eq_appendix:S_j} we have $\eta_1=1/\sqrt{5}$, $\eta_2=-1/\sqrt{5}$, $\varphi_1=(1+\sqrt{5})/2$ and $\varphi_2=(1-\sqrt{5})/2$. The initial condition ${\bf S}_0^{(N)}$ is specific for each entry in the transformation matrix. 

This formalism reduces the problem to diagonalize the recursion matrix $R^{(N)}$, i.e., to find the roots $\varphi_l$ of its characteristic polynomial; afterwards we must find the coefficients $\eta_{l}^{}$ in Eq.~\eqref{eq:Snb}.  

The first task is addressed by different methods depending on the chosen framework. For the open tight-binding system (Eq.~\eqref{eq_appendix:H_open}), in the geometrical study of  \emph{golden fields} it was found that diagonals of regular $N$-side polygons are proportional to the ratio of successive terms in generalized Fibonacci sequences \cite{golden_fields}. This correspondence, described by the \emph{diagonal product formula} (DPF), shows that characteristic polynomials can be expressed in terms of the Fibonacci polynomials of second kind $K_n(x)$ \cite{golden}, which in turn relate to the Chebyshev polynomials of second kind $U_N(x)$ \cite{golden}
\begin{equation}\label{eq:Px}
 K_{N+1}(x)=U_N(x/2)\,.
\end{equation}
Chebysev polynomials are a key element in our search for a general transformation matrix, inasmuch as is known that the roots of $U_N(x)$ are given by $x_k=\cos(k\pi/[N+1])$ \cite{mason2002chebyshev}, for $k=1,\ldots,N$. Substituting in Eq.~\eqref{eq:Px}, the roots of the $K_N(x)$ are obtained as
\begin{equation}\label{eq:vphi_appendix}
 \varphi_k=2\cos\left(\frac{k\pi}{N+1}\right)\,,\qquad\quad k=1,\ldots,N\,.
\end{equation}
The above derivation not only gives the eigenvalues of the recursive matrix $R$, but also illustrates the relation between a $N$-mode lattice, the Fibonacci-like sequences and the associated polynomial families.

To determine the $\eta_l$ coefficients, it is essential to examine the initial condition ${\bf S}_0$ of the succession behind the $F_{mnk}$. The direct manner to come by them is to look at the first terms in the expansion of Eq.~\eqref{eq_appendix:App_BCH}. This procedure must be carry out with caution since \textit{a priori} $r$ is not known neither how many initial conditions exist. In the case $N=6$, we found that $Q_1=1$, $Q_2=2$, $Q_3=-1$ and three initial conditions
\begin{align}\label{eq_appendix:S0_N=6}
    {\bf S}_0^{(6)} &=
    \begin{pmatrix}
        2\\
        0\\
        1
    \end{pmatrix},\quad
    \begin{pmatrix}
    1 \\
    0 \\
    0
    \end{pmatrix},\quad
    \begin{pmatrix}
    2 \\
    1 \\
    1
    \end{pmatrix}\,.
\end{align}
From Eq.~\eqref{eq_appendix:S0_N=6} all the sequences to describe $A^{(6)}$ were determined, namely
\begin{align}
    &\begin{pmatrix}
        2\\
        0\\
        1
    \end{pmatrix}\longrightarrow \left\{1,\ 0,\ 2,\ 1,\ 5,\ 5,\ 14,\ 19,\ 42,\ 66,\ 131,\cdots\right\}\,,\\
    &\begin{pmatrix}
    1 \\
    0 \\
    0
    \end{pmatrix}\longrightarrow \left\{0,\ 0,\ 1,\ 1,\ 3,\ 4,\ 9,\ 14,\ 28,\ 47,\ 89,\ 155,\cdots\right\}\,,\\
    &\begin{pmatrix}
    2 \\
    1 \\
    1
    \end{pmatrix}\longrightarrow \left\{1,\ 1,\ 2,\ 3,\ 6,\ 10,\ 19,\ 33,\ 61,\ 108,\ 197,\cdots\right\}\,.
\end{align}

Note that the expansion of Eq.~\eqref{eq_appendix:App_BCH} by means of the BCH formula can always be done regardless the complexity of the tight-binding interactions in the Hamiltonian. Notwithstanding, finding a formula to describe the integers sequences is not a straightforward task if no information about them is available in the OEIS. Given the recursive nature of the BCH formula, is expected that the expansion of \eqref{eq_appendix:App_BCH} give rise to recursive sequences. However, this is not always the case as we will see in the close $N$-mer.

A more manageable scenery comes to light when we faced the case of the closed array (Fig. \ref{fg:modelo} (c)), owing to the sequences appearing are just binomial coefficient. For example, for $N=5$ the entry $\bar A^{(5)}_{1,2}$ results in 
\begin{align}
    \nonumber \overline{A}_{1,2}^{(5)} &= e^{i(\delta-\pi/2)}\Bigg(Ct-\frac{3}{3!}\left(Ct\right)^3+\frac{10}{5!}\left(Ct\right)^{5}-\frac{35}{7!}\left(Ct\right)^{7}\\
     \nonumber&+\frac{126}{9!}\left(Ct\right)^{9}-\frac{462}{11!}\left(Ct\right)^{11}+\frac{1716}{13!}\left(Ct\right)^{13}-\cdots\Bigg)\\
     \nonumber&+e^{-4i(\delta-\pi/2)}\Bigg(\frac{\left(Ct\right)^4}{4!}-\frac{6}{6!}\left(Ct\right)^6+\frac{28}{8!}\left(Ct\right)^{8}\\
     \nonumber&-\frac{120}{10!}\left(Ct\right)^{10}
     +\frac{495}{12!}\left(Ct\right)^{12}-\frac{2002}{14!}\left(Ct\right)^{14}+\cdots\Bigg)\\
     \nonumber&+e^{6i(\delta-\pi/2)}\Bigg(\frac{\left(Ct\right)^{6}}{6!}-\frac{8}{8!}\left(Ct\right)^{8}+\frac{45}{10!}\left(Ct\right)^{10}\\
     \nonumber&-\frac{220}{12!}\left(Ct\right)^{12}
     +\frac{1001}{14!}\left(Ct\right)^{14}-\frac{4368}{16!}\left(Ct\right)^{16}+\cdots\Bigg)\\
     \nonumber&+e^{-9i(\delta-\pi/2)}\Bigg(-\frac{\left(Ct\right)^{9}}{9!}+\frac{11}{11!}\left(Ct\right)^{11}-\frac{78}{13!}\left(Ct\right)^{13}\\
     \nonumber&+\frac{445}{15!}\left(Ct\right)^{15}-\frac{2380}{17!}\left(Ct\right)^{17}+\frac{11628}{19!}\left(Ct\right)^{19}-\cdots\Bigg)\\
     \nonumber&+e^{11i(\delta-\pi/2)}\Bigg(\frac{\left(Ct\right)^{11}}{11!}-\frac{13}{13!}\left(Ct\right)^{13}+\frac{105}{15!}\left(Ct\right)^{15}\\
     \nonumber&-\frac{680}{17!}\left(Ct\right)^{17}+\frac{3876}{19!}\left(Ct\right)^{19}-\frac{20349}{21!}\left(Ct\right)^{21}+\cdots\Bigg)\\
     \nonumber&+e^{-14i(\delta-\pi/2)}\Bigg(\frac{\left(Ct\right)^{14}}{14!}-\frac{16}{16!}\left(Ct\right)^{16}+\frac{153}{18!}\left(Ct\right)^{18} \\
     \nonumber&-\frac{1140}{20!}\left(Ct\right)^{20}+\frac{7315}{22!}\left(Ct\right)^{22}-\frac{42504}{24!}\left(Ct\right)^{24}+\cdots\Bigg)\\
     \nonumber&+e^{16i(\delta-\pi/2)}\Bigg(\frac{\left(Ct\right)^{16}}{16!}-\frac{18}{18!}\left(Ct\right)^{18}+\frac{190}{20!}\left(Ct\right)^{20}\\
     \nonumber&-\frac{1540}{22!}\left(Ct\right)^{22}+\frac{10626}{24!}\left(Ct\right)^{24}-\frac{65780}{26!}\left(Ct\right)^{26}+\cdots\Bigg) \\
     \nonumber&+e^{-19i(\delta-\pi/2)}\Bigg(-\frac{\left(Ct\right)^{19}}{19!}+\frac{21}{21!}\left(Ct\right)^{21}-\frac{253}{23!}\left(Ct\right)^{23}\\
     \nonumber&+\frac{2300}{25!}\left(Ct\right)^{25}-\frac{17550}{27!}\left(Ct\right)^{27}+\frac{118755}{29!}\left(Ct\right)^{29}-\cdots\Bigg)\\
    \label{eq_appendix:closed_N=5}&+\cdots
\end{align}
The coefficients in the series Eq.~\eqref{eq_appendix:closed_N=5} with factor $e^{i(\delta-\pi/2)}$ where identified to be
\begin{align}\label{eq_appendix:binomial_coefficient}
    1,\ 3,\ 10,\ 35,\ 126,\ 462,\ 1716,\ \ldots \longrightarrow
    \begin{pmatrix}
        2n+1\\
        n+1
    \end{pmatrix}\,.
\end{align}
With these coefficients we were able to converge this series, obtaining
\begin{align}
    \nonumber \beta_{2,1}^{(5)}&=\sum_{n=0}^\infty
     \begin{pmatrix}
        2n+1\\
        n+1
    \end{pmatrix}
    \frac{(-1)^n}{(2n+1)!}\left(Ct\right)^{2n+1}\,, \\
    \nonumber \beta_{2,1}^{(5)}&= \sum_{n=0}^\infty \frac{(2n+1)!}{(n+1)!n!}\frac{(-1)^n}{(2n+1)!}\left(Ct\right)^{2n+1}\,,\\
    \nonumber \beta_{2,1}^{(5)}&= \sum_{n=0}^\infty\frac{(-1)^n}{n!(n+1)!}\left(Ct\right)^{2n+1}\,,\\
    \beta_{2,1}^{(5)} &= J_1(2Ct)\,.
\end{align}

Following the same procedure for the remaining complex exponentials, the $\beta_{2,b_l}^{(5)}$ factors were also turned to be Bessel functions of the first kind whose indexes were the $b_l$. This is summarized as follow
\begin{align}\label{eq:beta_close}
    \beta_{nb_l}^{(N)} &= J_{b_l}(2Ct)\,,\\
    b_l(5,2) &= \left\{1,\ 4,\ 6,\ 9,\ 11,\ 14,\ 16,\ 19,\cdots \right\}.
\end{align}

Replacing the series in Eq.~\eqref{eq_appendix:closed_N=5} by their respective Bessel functions results in 
\begin{align}\label{eq_appendix:A12_expansion}
   \nonumber \overline{A}_{1,2}^{(5)} &= e^{i(\delta-\pi/2)}J_1(2Ct)+e^{-4i(\delta-\pi/2)}J_4(2Ct)\\
   \nonumber &+e^{6i(\delta-\pi/2)}J_6(2Ct)
    -e^{-9i(\delta-\pi/2)}J_9(2Ct)\\
    \nonumber &+e^{11i(\delta-\pi/2)}J_{11}(2Ct)+e^{-14i(\delta-\pi/2)}J_{14}(2Ct)\\
   &+e^{16i(\delta-\pi/2)}J_{16}(2Ct)-e^{-19i(\delta-\pi/2)}J_{19}(2Ct)+\cdots\,.
\end{align}

The alternating minus sign over the terms of the series Eq.~\eqref{eq_appendix:A12_expansion} does not follow a pattern $(-1)^k$ like the minus sign over the argument of the complex exponentials. The rule for this minus sign is captured by the $\sigma_{
N,n,l}$ function defined in Eq.~\eqref{eq:sigma_sign}. Following the same strategy all the entries of $\bar A^{(N)}$ were found, obtaining
\begin{equation}\label{eq_appendix:A_1n}
     \begin{split}
         \overline{A}_{1,n}^{(N)} =
         \begin{cases}
         \alpha_{N}^{(1)}\; &N\ \text{odd},\ n=1\\
         \alpha_{N,n}^{(2)}\; &N\ \text{odd},\ n\leq(N+1)/2\\
         \mathcal{P}_c\alpha_{N,N-n+2}^{(2)} \; &N\ \text{odd},\ n>(N+1)/2\\
         \alpha_{N}^{(0)}\; &N\ \text{even},\ n=1\\
         \alpha_{N,n}^{(2)}\; &N\ \text{even},\ n\leq N/2\\
         \alpha_{N,n}^{(3)} \; &N\ \text{even},\ n=\frac{N}{2}+1,\ n \text{ odd}\\
         \alpha_{N,n}^{(4)} \; &N\ \text{even},\ n=\frac{N}{2}+1,\ n \text{ even}\\
         \alpha_{N,N-n+2}^{(2)\ast}\;  &N\ \text{even},\ n>\frac{N}{2}+1,\ n\text{ odd}\\
         -\alpha_{N,N-n+2}^{(2)\ast}\; &N\ \text{even},\ n>\frac{N}{2}+1,\ n\text{ even}
         \end{cases}
     \end{split}
 \end{equation}
where the $\alpha_{N,n}$ functions are given by
 \begin{align}\label{eq:alpha_N}
    \nonumber \alpha_{N}^{(0)} &= -J_0(2Ct)+2\sum_{l=0}^{\infty}
         \cos{(lN(\delta-\pi/2))}J_{lN}(2Ct)\,, \\  
    \nonumber \alpha_{N}^{(1)} &= -J_0(2Ct)\\
    \nonumber &+2\sum_{l=0}^{\infty}\Big(\cos{(2lN(\delta-\pi/2))}J_{2lN}(2Ct)\\
    \nonumber &+i\sin{((2l+1)N(\delta-\pi/2))}J_{(2l+1)N}(2Ct)\Big)\,,\\
    \nonumber \alpha_{N,n}^{(2)} &= \sum_{l=1}^{\infty}\sigma_{N,n,l}\exp{\left\{(-1)^{l+1}b_l(N,n)(\delta-\pi/2) i\right\}}\\
    \nonumber &\times J_{b_l}(2Ct)\,,\\
    \nonumber \alpha_{N,n}^{(3)} &= 2\sum_{l=0}^{\infty} \cos{\left(\frac{N}{2}(2l+1)(\delta-\pi/2)\right)}J_{N(2l+1)/2}(2Ct)\,, \\
    \alpha_{N,n}^{(4)} &= 2i\sum_{l=0}^{\infty} \sin{\left(\frac{N}{2}(2l+1)(\delta-\pi/2)\right)}J_{N(2l+1)/2}(2Ct)\,,
 \end{align}
with indexes $b_l(N,n)$ defined by the set 
\begin{equation}\label{eq:b_l}
    b_l(N,n) = \{k\in\mathbb{N}\ |\ k= n-1 \ \vee\  N-n+1\ {\rm mod}(N) \}\,,
\end{equation}
and the special sign function
\begin{equation}\label{eq:sigma_sign}
\begin{split}
    \sigma_{N,n,l} &=
    \begin{cases}
    \sigma_{1}(b_l(N,n))\quad  N \text{ odd } ,\ n \text{ even} \\
    \sigma_{2}(b_l(N,n)) \quad  N \text{ odd },\ n \text{ odd}\\
    (-1)^{l+1} \qquad\quad\! N \text{ even } ,\ n \text{ even} \\
    \quad\, 1 \quad\qquad\qquad\!\!\! N \text{ even } ,\ n \text{ odd}
    \end{cases},\\
    \sigma_{1}(b_l(N,n)) &= 
        \begin{cases}
            -1\quad \ b_l(N,n)=2N-n+1 \ {\rm mod}(2N)\\
            \;\;\; 1\quad \text{Otherwise}
        \end{cases}\\
        \sigma_{2}(b_l(N,n)) &= 
        \begin{cases},
            -1\quad \ b_l(N,n)=N-n+1 \ {\rm mod}(N)\\
            \;\;\; 1\quad \text{Otherwise}
        \end{cases},
         \end{split}
\end{equation}
with a parity conjugation operator
\begin{equation}
    \mathcal{P}_c e^{i n\delta}=
    \begin{cases}
    -e^{-i n\delta}\quad n \text{ odd}\\
    \quad\! e^{-i n\delta} \quad n \text{ even}\,.
    \end{cases}
\end{equation}
    The $\alpha_{N,n}$ functions for the entries of $A^{(N)}$ resembles the Jacobi-Anger expansions but with a major detail on the indexes of the series. This feature make us unable to converge the series in Eq.~\eqref{eq:alpha_N}.
    It is important to highlight the versatility of the previous formalism, since Eqs.~\eqref{eq_appendix:A_mn} and \eqref{eq:Snb} hold for any set of coupling coefficients, in addition to the two one-dimensional cases studied here. This allows us to extend our study to any two-dimensional system that can be mapped to a one-dimensional matrix, provided that the appropriate integer sequence (or its recurrence matrix) is known.

\subsection*{d. Path method}
\label{Appendix:Quantum random walks and graph theory}

Coefficients $\eta_{j}$ in Eq.~\eqref{eq:Snb} are still required to find the sequence describing the entries of the transformation matrix. However, we can use the link between quantum random walks and graph theory, provided by the aforementioned equivalence between the coupling matrix $\mathcal C$ (corresponding to the single-particle representation of $H$) and the adjacency matrix $\mathcal{J}$ of the associated path graph $P_N$. If $G$ is a graph with weights at the edges and adjacency matrix $\mathcal{J}_{N,\varrho}$, where $N$ is the number of vertices and $\varrho$ is the weight of the loops, then the sum of weights of walks of length $m$ from vertex $i$ to $j$ is given by $e_i^T\cdot \mathcal{J}^m_{N,\varrho}\cdot e_j$. Defining $Z_{i,j}^N(m,\ell)$ as the number of walks from $i$ to $j$ of length $m$, using $\ell$ loops in $P_n$, the following can be obteined \cite{Znumbers}

\begin{equation}\label{eq:sumzeta}
    \sum_{\ell=0}^m Z_{i,j}^N(m,\ell)\varrho^\ell=e_i^T\cdot \mathcal{J}^m_{N,\varrho}\cdot e_j
\end{equation}
This is precisely the expression we need in order to write the Eq.\eqref{eq:App_commutator_n} more effectively. Obtaining an expression for the left-hand side of Eq.\eqref{eq:sumzeta} is straightforward when considering the spectrum of the adjacency matrix. If $\mathcal{J}_{N,\varrho}$ admits an orthonormal basis $(v_1,v_2,\ldots,v_N)$ of eigenvectors and $\lambda_i \in \mathbb{C}$ is the eigenvalue corresponding to $v_i=(v_{1,i},v_{2,i},\ldots,v_{N,i})$, then

\begin{equation}\label{eq:sumacaminos}
    e_i^T\cdot \mathcal{J}^m_{N,\varrho}\cdot e_j=\sum_{k=1}^{N}\lambda_{k}^{m}v_{i,k}v_{j,k}.
\end{equation}

In the particular case of $\mathcal{J}_{N,0}$ the eigenvalues $\lambda_k$ and eigenvectors $v_k$ are given by 

\begin{equation}\label{eq:autovalorad}
    \lambda_k= 2\cos\left(\frac{k\pi}{N+1}\right),
\end{equation}

and

\begin{equation}\label{eq:autovectorad}
\begin{aligned} 
    v_k = \biggl(&\sin\left(1\cdot\frac{k\pi}{N+1}\right), 
    \sin\left(2\cdot\frac{k\pi}{N+1}\right),\sin\left(3\cdot\frac{k\pi}{N+1}\right), \\
    &\ldots, 
    N
    \cdot\sin\left(\frac{k\pi}{N+1}\right)\biggr).
\end{aligned}
\end{equation}
for $k=1,2,\ldots,N$. Note that the Eq. \eqref{eq:autovalorad} is identical to the expression obtained in Eq. \eqref{eq:vphi_appendix}. This illustrates a close relationship between the Chebyshev polynomials of the second kind and their roots and eigenvalues of the adjacency matrix displayed above. Replacing Eqs. \eqref{eq:autovalorad} and \eqref{eq:autovectorad} in Eq. \eqref{eq:sumacaminos} and using Eq.\eqref{eq:sumzeta} enables us to obtain an explicit expression for the numbers $Z_{i,j}^N(m,\ell)$ in terms of trigonometric sums. Assuming $\ell=0$, in the case we have open arrays the $Z_{i,j}^N(m)$ numbers are given by 

\begin{equation}\label{eq:Zn}
 \begin{split}
 Z^{(N_{\rm{open}})}_{i,j}(m)=\frac{2}{N+1}&\sum\limits_{k=1}^N\left(2\cos\left(\frac{k\pi}{N+1}\right)\right)^m\\
 &\times\sin\left(\frac{ik\pi}{N+1}\right)\sin\left(\frac{jk\pi}{N+1}\right),
 \end{split}
 \end{equation}
for an $N$-modes open array with first-neighbors coupling, and for cycle graphs $C_n$, which correspond to paths where the walk begins at vertex $i$ and ends at vertex $j$, and which we associate with closed arrays, the numbers are given by
 \begin{equation}\label{qe:wn}
     \begin{split}
     Z^{(N_{\rm{closed}})}_{i,j}(m)&=\frac{1}{n}\sum_{k=0}^{n-1}\left(2\cos{\left(k\frac{2\pi}{n}\right)}\right)^m\cos{\left(k\frac{2\pi (i-j)}{n}\right)}\,\\
     n &= \begin{cases}
         N\quad \phantom{2}N\ {\rm even}\\
         2N\quad N\ {\rm odd}
     \end{cases}
     \end{split}
 \end{equation}


Equations \eqref{eq:Snb} and \eqref{eq:Zn} show the formal correspondence between the $m$-th term in the series expansion of $A_{i,j}^{(N)}$ and the total number of all $m$-steps paths, i.e., the path length corresponds to the order of the terms in the approximation given by Eq.~\eqref{eq_appendix:A_mn}. This enables a straightforward physical interpretation of the path count: while adding $Z_{i,j}^{(N)}(m)$ for $0<m<\infty$ would give the number of all possible paths between vertices $i$ and $j$, $A_{i,j}^{(N)}$, on the other hand, captures the superposition of all the contributions resulting in the excitation of waveguide $a_i$ by an initial excitation of waveguide $a_j$.  Indeed, the $Z$ numbers can be considered a universal quantity, and terms in many numerical successions, including the Fibonacci and Catalan sequences, can be obtained as particular cases of them \cite{Znumbers}.

In order to construct the complete entry $A_{\mu,\nu}^{(N)}$, we must sum the number $Z_{i,j}^{(N)}(m)$ defined in Eq. \eqref{eq:Zn} over all the path lengths $m$ from 0 to $\infty$. Also, we need to modify the sum in the equation to take into account the complex phases $\delta_i$ present in the coupling coefficients $\Lambda_i$.
By doing so, we finally obtain
\begin{equation}\label{eq:Amn_appendix}
 \begin{split}
 A_{\mu,\nu}^{(N)}= \frac{2}{N+1}\sum_{k=1}^N \exp\left(-2i\cos\left[\frac{k\pi}{N+1}\right]C t\right)S_{\mu\nu k}\,,
 \end{split}
\end{equation}
where
\begin{equation}\label{eq:Smnk}
\begin{split}
  S_{\mu\nu k}=&\exp\left(-i\left[\frac{\pi}{2}(\mu-\nu)+\Delta_{\mu-1}-\Delta_{\nu-1}\right]\right)\\
 &\times \sin\left(\frac{\mu k\pi}{N+1}\right)\sin\left(\frac{\nu k\pi}{N+1}\right)
\end{split}
\end{equation}
and the terms $\Delta_\nu=\delta_1+\delta_2+\cdots+\delta_\nu$ contain the information about the phases of the coupling coefficients $C_j=C\exp(i\delta_j)$, $C$ being their real amplitude. 
On the other hand, the \emph{closed} system, i.e. the $N$-mer where the first and the $N$-th mode are also coupled, is described by a transformation matrix $\overline{A}^{(N)}$, whose $n$-th row is obtained by \emph{cycling} $n$ times the first row with entries
\begin{equation}
     \begin{split}
         \overline{A}_{1,n}^{(N)} = \frac{e^{2iC t}}{N}&+\frac{2}{N}\sum_{k=1}^{(N-1)/2}(-1)^{(n-1)k}\cos{\left((n-1)k\frac{\pi}{N}\right)}\\
         &\times\exp{\left( (-1)^{k}2iC t \cos{\left(k \frac{\pi}{N}\right)}\right)}\,
     \end{split}
 \end{equation}
for $N$ odd. For $N$ even its elements are given by
 \begin{equation}
     \begin{split}
         \overline{A}_{1,n}^{(N)} =& \frac{1}{N} \sum_{k=0}^{N-1}\cos{\Big((n-1)k\text{\small $\frac{2\pi}{N}$}\Big)}\\
         &\times
         \begin{cases}
         \cos{\Big(2C t\cos{\Big(\text{\small $k\frac{2\pi}{N}$}\Big)}\Big)} \hspace{0.4cm}\text{$n$ odd}\\
         i\sin{\Big(2C t\cos{\Big(\text{\small $k\frac{2\pi}{N}$}\Big)}\Big)} \hspace{0.4cm}\text{$n$ even}
    \end{cases}\,.
     \end{split}
 \end{equation}



\bibliography{nmero}
\bibliographystyle{unsrt}



\end{document}